%% file: main.tex
\documentclass[iop]{emulateapj}
\usepackage{verbatim}
\usepackage{mathtools}
\usepackage{bm}        
\usepackage{amssymb}   
\usepackage{booktabs}
\usepackage{subfigure}
\usepackage{graphicx}
\usepackage{amsmath,amssymb}
\RequirePackage{import}
\usepackage{import}
\usepackage{appendix}
\usepackage{xcolor}
\usepackage{comment}

\usepackage{natbib}

\begin{document}

\title{Quantifying the projected suppression of cluster escape velocity profiles}

\author{Vitali Halenka\altaffilmark{1}, Christopher J. Miller\altaffilmark{1,2}, Paige Vansickle\altaffilmark{2}  } 

\altaffiltext{1}{Department of Physics, University of Michigan, Ann Arbor, MI 48109 USA}
\altaffiltext{2}{Department of Astronomy, University of Michigan, Ann Arbor, MI 48109, USA}
\email{vithal@umich.edu}

\begin{abstract}

    The 3D radial escape-velocity profile of galaxy clusters has been suggested to be a promising and competitive tool for constraining mass profiles and cosmological parameters in an accelerating universe. However, the observed line-of-sight escape profile is known to be suppressed compared to the underlying 3D radial (or tangential) escape profile. Past work has suggested that velocity anisotropy in the phase-space data is the root cause. 
    Instead, we find that the observed suppression is from the statistical undersampling of the phase spaces and that the 3D radial escape edge can be accurately inferred from projected data. 
    We build an analytical model for this suppression that only requires the number of observed galaxies $N$ in the phase-space data within the sky-projected range $0.3 \le r_\perp/R_{200, \text{critical}} \le 1$. The radially averaged suppression function is an inverse power law $\langle Z_\text{v} \rangle =  1 +  (N_0/N)^\lambda$ with $N_0 = 17.818$ and $\lambda= 0.362$. 
    We test our model with $N$-body simulations, using dark matter particles, subhalos, and semianalytic galaxies as the phase-space tracers, and find excellent agreement. We also assess the model for systematic biases from cosmology ($\Omega_{\Lambda}$, $H_0$), cluster mass ($M_{200, \text{critical}}$), and velocity anisotropy ($\beta$). We find that varying these parameters over large ranges can impart a maximal additional fractional change in $\langle Z_\text{v} \rangle$ of $2.7\%$. These systematics are highly subdominant (by at least a factor of 13.7) to the suppression from $N$.

\end{abstract}

\section{Introduction}\label{intro}
\input{intro.tex}

\section{Motivation}\label{section:motiv}
\input{motivation.tex}

\section{Line-of-sight Velocities and Escape Speed}\label{section:approach1}
\input{approach1.tex}

\section{AGAMA-based Phase Spaces}\label{approach_desription}
\input{approach.tex}

\section{Results}\label{results}
\input{results.tex}

\section{Summary}\label{disc_concl}
\input{disc_concl.tex}

\section{Acknowledgments}
We thank August E. Evrard, Dragan Huterer, and Mario Mateo for useful discussions. This material is based upon work supported by the National Science Foundation under grant No. 1812739. 

\bibliographystyle{apj}
\bibliography{main.bib}

\end{document}

%% file: intro.tex
Galaxy clusters are the largest most recently formed cosmological objects. Galaxies inside the potential are sparsely distributed and represent a small fraction of the baryonic content. The majority of the baryons in clusters are in the mostly smooth gaseous intracluster medium. In the current $\Lambda$CDM paradigm, the cluster potential is dominated by dark matter, which, except gravitationally, is not known to interact with the baryons. Through the Poisson equation, the cluster potential governs the dynamics of all massive tracers in the cluster, including the galaxies. In this scenario, we expect tracers on elliptical orbits to have been accelerated to escape speeds at their closest approach and that these tracers will be largely unaffected by dynamical friction, tidal interactions, or encounters with other tracers (see \cite{Aguilar2008} for a review). At any given radius away from the cluster center, there will be tracers that are moving at the escape speed. Therefore, the escape-velocity profile becomes a property of clusters representing the underlying potential with few astrophysical systematic issues \citep{Miller:2016fku}. 

The escape-velocity profile, $v_\text{esc}(r)$, of a cluster is a clearly defined edge in the radius/velocity phase-space diagram. Only the tracers with the maximum possible radial or tangential 1D speed will contribute to this edge \citep{Behroozi2013}. The power of utilizing the observed $v_\text{esc}(r)$ is in its direct connection to the total potential, enabling cluster-mass estimations and tests of gravity on the largest scales in the weak-field limit and placing constraints on the $\Lambda$CDM cosmological parameters \citep{Gifford2013a, Gifford:2013ufa,Stark:2016mrr,Stark2017}. 

Up until now, simulations have always shown that the observed edge is lower than the underlying radial or tangential $v_\text{esc}$ profile. Because of this, most mass profile modelers using caustics have utilized $N$-body simulations to calibrate the amount of suppression in the projected escape-velocity profile \citep{Diaferio:1997mq, Diaferio:1999wg, Serra2011, Gifford:2013ufa}. However, \citet{Stark:2016dxf} used a novel technique where they combined weak-lensing mass profiles and cluster phase-space data to observationally constrain the suppression without simulations. Combined, these studies find that the projected edge is about $60-80\%$ suppressed in comparison with the 3D radial escape edge. This is the dominant systematic when using the observed phase-space edge to infer cluster-mass profiles or in cosmological parameter estimation.

In this work, we take a new approach to determine the amount of projected escape-edge suppression, which does not require simulations or weak-lensing observations. Our approach is rather simple and is based on populating mock halos with galaxies on realistic orbits. While these mock phase spaces do not contain the full dynamical information of a true massive and fully evolved halo, we show that the 3D radial and projected phase-space edges closely match those of evolved cosmological $N$-body simulations.

The plan of the paper is following. We start with Sections 2 and 3, where we review the connection between the escape-velocity profile, the gravitational potential, and cosmology as motivation for understanding the suppression of the projected escape profile. In Section 4 we develop an analytical approach to model the escape profile of cluster phase spaces. In Section 5 we apply our model to mock cluster samples and in $N$-body simulations. We finish with a summary and discussion.

Throughout the paper and where necessary, we use a flat standard cosmology with $\Omega_\text{M} = 0.3$ and $\Omega_{\Lambda} = 1 - \Omega_\text{M}$, and  $H_0 = 100h \, \text{km} \, \text{s}^{-1} \, \text{Mpc}^{-1}$ with $h = 0.7$ is assumed. We refer to the following quantities $R_{200}$ and $M_{200}$ as the radius and the mass of clusters at the point when the cumulative interior density drops to $200 \rho_\text{c,z}$, where $\rho_\text{c,z} = 3H^2 / (8 \pi G)$ is the critical density of the universe at redshift $z$ and $E(z) = H(z)/H_0 = \sqrt{\Omega_\Lambda + \Omega_M (1+z)^3}$. The connection between $R_{200}$ and $M_{200}$ for spherical systems is by definition $M_{200} = \frac{4\pi}{3} (200 \rho_\text{c,z}) R^3_{200}$.

%% file: motivation.tex
\subsection{Escape-velocity Profile in an Expanding Universe}\label{theory}

The main conclusion of general relativity is the Einstein equation, which relates matter/energy density to the curvature of space-time \citep{Einstein1916, Jacobson1995}. Through the Poisson equation, this curvature in turn governs the dynamical behavior of the local matter. \citet{Nandra2012a} derived an invariant fully general relativistic expression, valid for arbitrary spherically symmetric systems, for the force required to hold a test particle at rest relative to the central point mass in an accelerating universe. As then also noted by \citet{Behroozi2013}, in a $\Lambda$CDM universe there is a location in space ($r_\text{eq}$) that is well defined and relative to a massive body (like a cluster), where the radially inward gravitational force acting on a tracer from the massive object is equivalent to the effective radially outward force due to the acceleration of the underlying space-time,
\begin{equation}\label{r_eq}
    r_\text{eq} = \Bigr( -\frac{GM}{q(z)H^2(z)} \Bigr)^{1/3},
\end{equation}
where $G$ is the gravitational constant, $M$ is the mass of the cluster, $H(z)$ is the Hubble expansion parameter, and the deceleration parameter is $q(z) = \frac{1}{2} \Omega_\text{m}(z) - \Omega_\Lambda(z)$. In the flat standard cosmology, $r_\text{eq}$ is $\sim$$8$-$9$ times greater than $r_{200}$.

An important observational consequence of Equation (\ref{r_eq}) is in the definition of the escape velocity on cosmological scales. In the Newtonian or weak-field limit, the escape velocity is defined by the potential
\begin{equation}\label{v_esc}
    v_\text{esc} = \sqrt{-2\Phi},
\end{equation}
where $\Phi$ is the total potential that includes the gravitational potential ($\phi$) as well as the potential in the expanding space-time \citep{Riess:1998cb, Calder2008}.
As discussed in \cite{Behroozi2013}, the 3D radial\footnote{Objects on tangential escape trajectories require slightly more energy to escape than those on radial orbits as presented in \citet{Behroozi2013}. However, the difference is small inside the virialized region.} escape-velocity profile is of the following form
\begin{equation}\label{v_esc_full}
    v_\text{esc} = \sqrt{-2[\phi(r) - \phi(r_\text{eq})] - q(z)H^2(z)[r^2 - r_\text{eq}^2]}.
\end{equation}
Equation (\ref{v_esc_full}) tells us that the slope of the escape velocity profile runs downward with radius due to the $q(z)H^2(z)r^2$ contribution and also that the overall amplitude of the escape edge shifts downward due to $r_\text{eq}$, the latter being the dominant effect. Equation (\ref{v_esc_full}) was tested to high precision and accuracy (percent level) using $N$-body simulations \citep{Miller:2016fku}.

We can make an observation of the escape-velocity profile of a cluster in projection on the sky. Likewise, we can measure the gravitational potential profile $\phi(r)$ from the gravitationally lensed shear of the background galaxies. Combined, such data make a powerful cosmological probe \citep{Stark2017}. The issue we address in this paper is the statistical effect of undersampled phase spaces, which leads to a suppression of the underlying escape-velocity profile.

\subsection{Observed Galaxy Cluster  Radius/Velocity Phase Spaces}\label{connect_theory_data}
\begin{figure}[t]
\centering
\includegraphics[width=0.99\linewidth]{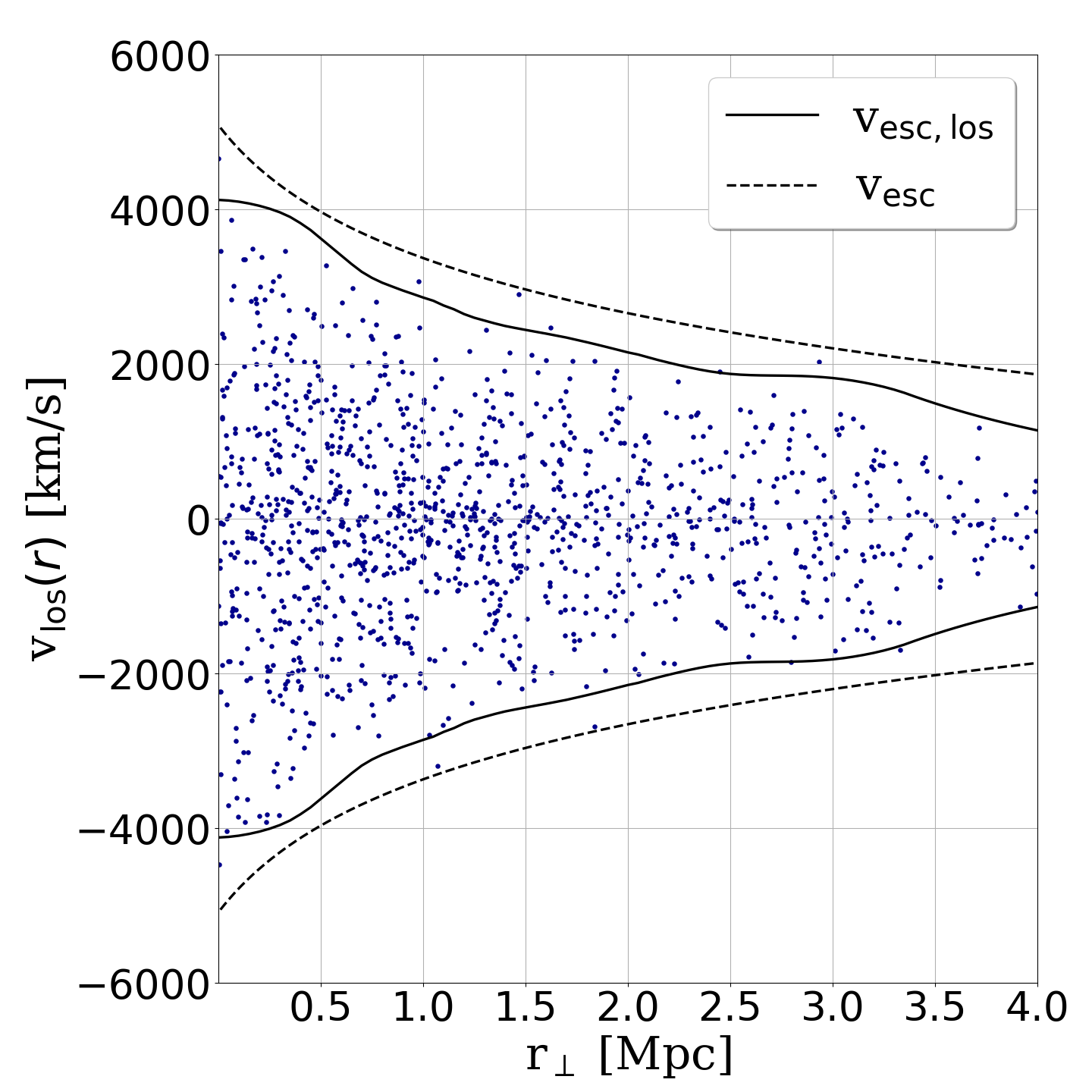}
\caption{An example projected phase space, i.e. line-of-sight velocity $v_\text{los}$ [km/s] vs. radial distance $r_\perp$ [Mpc] away from the center of a galaxy cluster. Dots correspond to positions and velocities of individual galaxies. Dashed black lines correspond to a three-dimensional radial escape-velocity profile inferred from this cluster's mass profile using weak-lensing measurements and a standard $\Lambda$CDM cosmology for Equation (\ref{v_esc_full}). Solid black lines correspond to the maximum observed on the projected phase-space diagram velocity profile measured by using an interloper removal prescription proposed by \cite{Gifford:2013ufa}. This paper aims to explain the difference between the amplitudes of the weak-lensing inferred and observed escape profiles. 
}
\label{fig:esc_prof_individual}
\end{figure}

We acquire galaxy velocities along the line of sight ($v_\text{los}$) by measuring their redshifts ($z_\text{g}$) as well as the redshift of the cluster redshift center ($z_\text{c}$),

\begin{equation}\label{v_los}
    v_\text{los} = c \, \frac{z_\text{g}-z_\text{c}}{1+z_\text{c}},
\end{equation}
where $c$ is the speed of light.

We then infer the galaxy-projected radial distances from the center of the cluster ($r_\perp$) using a specified cosmology,
\begin{equation}\label{r_g}
    r_\perp = r_\theta \Bigr( \frac{1}{1+z_\text{c}}\frac{c}{H_0} \int_0^{z_\text{g}} \frac{dz'}{E(z')} \Bigr), 
\end{equation}
where $r_\theta$ and $r_\perp$ are the angular and radial separations between the galaxy and the center of the cluster\footnote{We assume that with a large-enough galaxy sample in the phase-space data ($\sim 100$ galaxies), or with ancillary X-ray data, the cluster center can be well determined. Clusters that show signs of mergers or other significant substructure can be excluded from this type of scientific analysis.}.  By knowing both ($v_\text{los}$) and ($r_\perp$) we create a projected phase space for each cluster, i.e. $v_\text{los}$ vs. $r_\perp$ (see an example in Fig. \ref{fig:esc_prof_individual}). The edge in the projected phase space is the maximum observed velocity profile $v_\text{esc,los}$ (see solid lines on Figure \ref{fig:esc_prof_individual}). 

\cite{Diaferio:1997mq} and \cite{Diaferio:1999wg} laid the initial foundations for the projected escape-velocity technique using the idea of ``caustics'' in the 2D phase-space density. They worked in potential units, such that they were using the maximum observed velocity to infer the square of the escape-velocity profile. Thus, the underlying premise involves a geometric projection of the classic anisotropy parameter, $\beta$. Formally, the velocity anisotropy is
\begin{equation}\label{beta_old}
    \beta = 1 - \frac{\sigma^2_\theta}{\sigma^2_\text{r}},
\end{equation}
where $\sigma_\theta$ and $\sigma_\text{r}$ are tangential and radial velocity dispersions. The dispersion is
\begin{equation}\label{f-la:dispersion}
    \sigma^2(r) = \langle v^2(r) \rangle,
\end{equation}
where the $v(r)$ are velocities of individual galaxies measured with respect to zero (i.e. to the cluster frame of reference) and the average $\langle \cdot \rangle$ is over all the galaxies inside a 3D radial bin at $r$ with a width $\Delta r$. Using geometric arguments, Diaferio posited the following relation between the line of sight. and 3D escape velocity of a cluster:
\begin{equation}
\begin{aligned}
\langle v^2_\text{esc,los}\rangle (r) & = \frac{1-\beta(r)}{3-2\beta(r)}\langle v_\text{esc}^2\rangle(r) \\
& = (g(\beta(r)))^{-1}\langle v_\text{esc}^2\rangle(r),
\end{aligned}
\label{eq:v_proj}
\end{equation}
where $g(\beta(r)) \equiv (3-2\beta(r)) / (1-\beta(r))$.

The above premise suffers from an important statistical issue that was never addressed. The problem lies in the fact that it is based on projected dispersions averaged over projected radii (see Figure \ref{fig:front_side_views}). The dispersion measured in the small box B is not the same as that of the dispersion measured through the integrated line of sight. By necessity of monotonic potentials (see Figure \ref{fig:3d_los_velocities}), the dispersions in boxes A and C must be smaller than those at B. By including tracers in boxes A and C as representative of the average dispersion in box B, one is necessarily biasing the result.

As another approach in assessing the validity of Equation (\ref{eq:v_proj}), consider a densely sampled phase space (e.g., of dark matter particles). With enough sampling, one would surely identify a tracer near the escape speed with its velocity perfectly aligned with the line of sight at a projected radius identical to the 3D radius (i.e., red arrow at the position K in Figure \ref{fig:front_side_views}). In this case, one could observe the full 3D escape speed at this radius regardless of the radially averaged anisotropy of the underlying system. Any tracer that is not at position K, but is still along the line of sight, must necessarily experience a lower potential and escape speed due to the monotonically decreasing potential (see Figure \ref{fig:3d_los_velocities}).

%% file: approach1.tex
\begin{figure}
\centering
\includegraphics[width=0.95\linewidth]{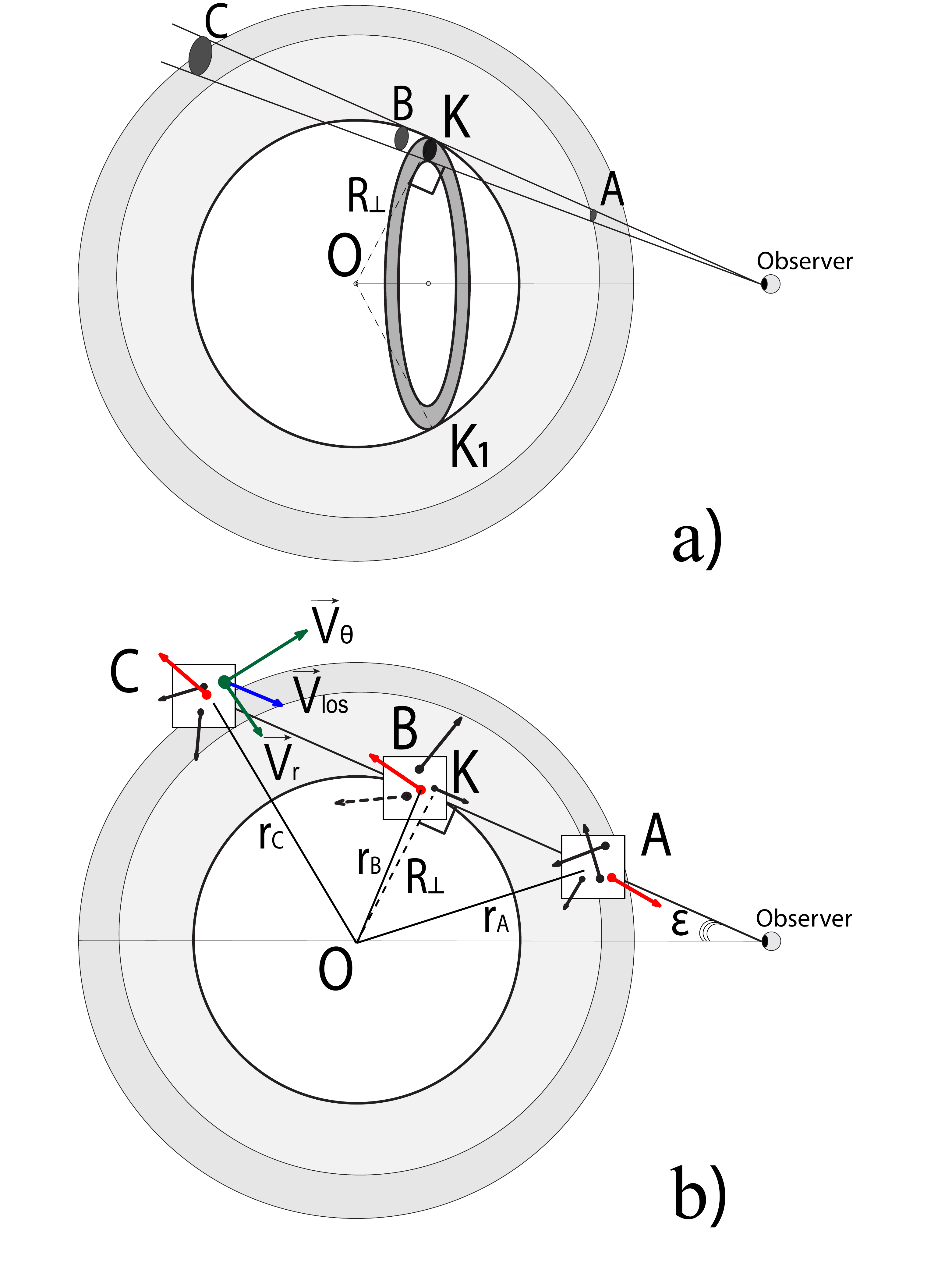}
\caption{This figure describes the geometry of our spherical systems. (a) While in reality the areas A, B, and C are spatially separated, for the outside observer they have the same position on the sky. The gray ring KK$_1$ represents the area that is equally separated from the center of the cluster O. Any galaxy in this ring as well as on the sphere KK$_1$ will be in the gray band $R_\perp$ on the three-dimensional phase space in Figure \ref{fig:3d_los_velocities}(a). All the galaxies in the cone that is created by circling the line of sight AC around the ring KK$_1$ will be in the gray band $R_\perp$ in Figure \ref{fig:3d_los_velocities}(b). 
(b) Arrows represent the velocities of individual galaxies. Black (red) arrows are the galaxies with velocity directions not aligned (aligned) with the line of sight AC. Any vector velocity of a galaxy (see Equation (\ref{v_tot})) is a sum of the tangential, radial (green arrows in the box C), and azimuthal (not presented due to direction pointing in/out of the plane of the figure) velocity components. The magnitude of the line-of-sight velocity (blue arrow in the box C) can be expressed in terms of tangential and radial components (see Equation (\ref{v_los_perp_rC})). The angle $\epsilon$ between the line of sight AC and the line that connects the center of the cluster O and the observer is much smaller in reality due to the distance from the observer to the cluster being much larger in comparison to the size of a cluster. The distances between different points: OC$=r_\text{C}$, OB$=r_\text{B}$, OK$=R_\perp$ and OA$=r_\text{A}$. OK $\perp$ AC.
}
\label{fig:front_side_views}
\end{figure}

\begin{figure}
\centering
\includegraphics[width=0.95\linewidth]{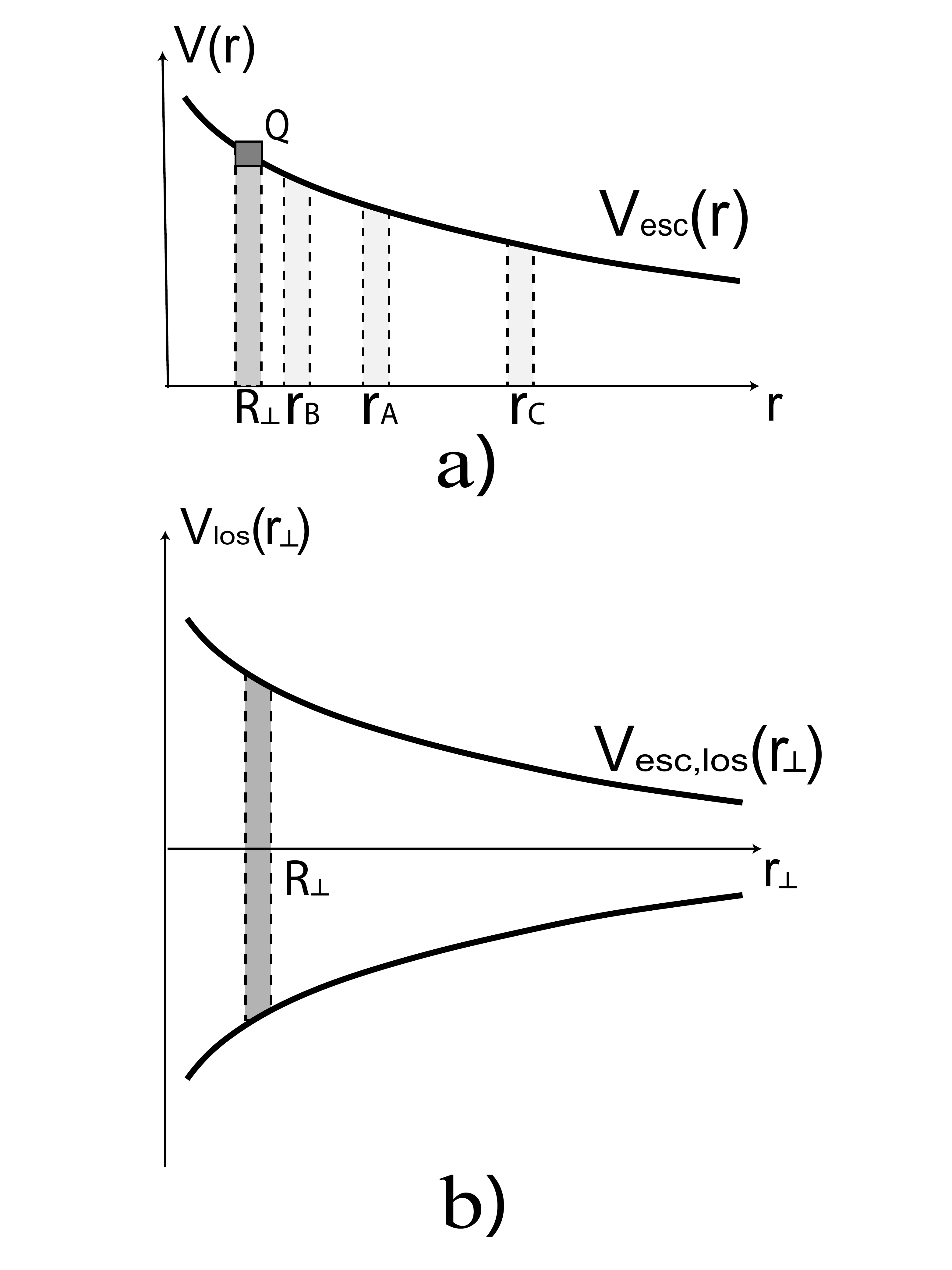}
\caption{A toy model of the phase-space edge for tracers in Figure \ref{fig:front_side_views}. (a) The phase-space envelope, i.e., the peculiar velocity (km/s) vs. distance $r$ (Mpc) away from the center of the cluster. The $v_\text{esc}(r)$ line is a measure of gravitational potential (see formula \ref{v_esc}). Gray bands $r_\text{B}, r_\text{A}$ and $r_\text{C}$ represent areas on the phase space where galaxies from dark small ellipses (Figure \ref{fig:front_side_views}(a)) and boxes (Figure \ref{fig:front_side_views}(b)) B, A, and C would be observed. Box Q represents the area, where all the galaxies with $v_\text{esc}(R_\perp)$ from the thin shell with radius $R_\perp$ and center O would be observed in the phase space. 
(b) Observed phase-space envelope, i.e., observed peculiar velocity (km/s) vs. radial distance $r_\perp$ (Mpc) away from the center of the cluster. The $v_\text{esc,los}(r_\perp)$ lines are the maximum observed velocities that can be obtained by taking the partial derivative $\partial v_\text{esc,los}(r, r_\perp) / \partial r = 0$. Similarly, the solid black lines in Figure \ref{fig:esc_prof_individual} are the observed maximum velocities. The gray band $R_\perp$ represents where galaxies from the ellipses (Figure \ref{fig:front_side_views}(a)) and the boxes (Figure \ref{fig:front_side_views}(b)) B, A, and C would be observed in the observed phase space. Note, while the phase space in (a) is always positive (presenting the absolute value of velocity relatively to the center of the cluster), the observed phase space can be negative as well due to galaxy velocities being able to point toward and away from the observer.
}
\label{fig:3d_los_velocities}
\end{figure}

\subsection{Relative Position}\label{subsection:relative_position)}

From the perspective of the distant observer, many cluster galaxies are at the same distance\footnote{In the full statistical analysis, we include interlopers that are projected into the cluster but lie well outside the virial radius.}. Some of the galaxies are physically closer to the observer (arrows in the box A in Figure \ref{fig:front_side_views}(b)), some farther away from the observer (box C), and some are somewhere at an intermediate distance (box B) such that the projected radius is close in value to the 3D radius. The 3D and projected phase-space radial locations of these boxes are shown in Figure \ref{fig:3d_los_velocities}. For the distant observer, the relative position of all of the boxes is equal to OK$=R_\perp$, a cone that is created by circling the line of sight AC around the ring KK$_1$.


\subsection{The Maximum Observed Velocity}
We next address the tracer-projected velocity in the context of its maximum because we are concerned with the maximum velocity at any radius (i.e. the escape speed).
The total velocity can be written down in terms of three individual vector components as
\begin{equation}\label{v_tot}
    \pmb{v}(r) = \pmb{v}_\theta (r) + \pmb{v}_\phi (r) + \pmb{v}_\text{r}(r),    
\end{equation}
where $\pmb{v}_\theta (r)$, $\pmb{v}_\phi (r)$, and $\pmb{v}_\text{r}(r)$ (see the green vectors in Figure \ref{fig:front_side_views}(b)) are the tangential, azimuthal, and radial components of the total velocity $\pmb{v}(r)$.

The projected component of $\pmb{v}(r)$ along the line of sight (see the blue vector in Figure \ref{fig:front_side_views}(b)) is
\begin{equation}\label{v_los_perp_rC}
    v_\text{los}(r_\text{C}) = v_\theta(r_\text{C})\cos({\frac{\pi}{2} - \psi}) - v_\text{r}(r_\text{C})\cos{\psi},
\end{equation}
where $\psi = \measuredangle \text{OCB}$ and $r_C$ is the actual distance between point C and the center of the cluster O. We can rewrite expression \ref{v_los_perp_rC} relative to the cluster center as
\begin{equation}\label{v_los_perp}
    v_\text{los}(r, r_\perp) = v_{\theta}(r)\frac{r_\perp}{r} - v_\text{r}(r)\frac{(r^2 - r^2_\perp)^{1/2}}{r},
\end{equation}
where $r_\text{C}$ ($R_\perp$) has been substituted by $r$ ($r_\perp$).

The maximum velocity $v_\text{esc,los}$ is what we actually observe as an edge in the phase space (see the solid lines in Figure \ref{fig:esc_prof_individual}), and it can be derived by solving the partial differential equation $\partial v_\text{esc,los}(r, r_\perp) / \partial r = 0$.    

The maximum observed velocity ($v_\text{esc,los}$) is a function of both $v_\text{r}$ and $v_\theta$. Because of the monotonic nature of the cluster potential (and escape) profiles, this maximum should only occur where $r=r_\text{max} =  r_\perp$. However, this would happen rarely because few galaxies have $r_\perp$ close to $r$ and have a velocity at the escape speed and aligned along the line of sight. In highly sampled systems, these rare alignments should happen often enough to accurately trace the 3D escape edge in projected coordinates. As the sampling becomes more sparse, these chance alignments become increasingly rare, thus suppressing the escape edge. We test this hypothesis in the following sections.



\subsection{Quantifying the Escape-velocity Suppression}
To quantify the escape-velocity suppression, we introduce the factor $Z_\text{v}$ by which the 3D radial escape velocity ($v_\text{esc}$) is suppressed in order to produce the observed maximum velocity $v_\text{esc,los}$
\begin{equation}\label{ratio_vel}
    Z_\text{v}(r_\perp) = \frac{v_\text{esc}(r_\perp)}{v_\text{esc,los}(r_\perp)} .
\end{equation}

%% file: approach.tex
Our statistical approach uses the Action-based Galaxy
Modeling Architecture (AGAMA) \citep{AGAMA} framework (see Section \ref{subsection:agama_implement} below) to forward model a cluster phase space that would mimic the basic characteristics of a predefined galaxy cluster (observed or simulated). There is one free parameter in the model that we later constrain, which is the suppression function $Z_\text{v}$. This parameter is not expressed analytically and must be calculated after the 3D phase spaces are projected onto the plane of the sky. 

We employ a statistical analysis called approximate Bayesian computation (ABC), which is designed for scenarios where a full analytical likelihood is not readily available. The goal of ABC is to develop a forward map and apply it with input parameters to simulate real observations, thus bypassing a direct calculation of a likelihood. The model parameters are drawn from some prior distribution. The simulated data are then reduced into a summary statistic. A posterior probability distribution is then approximated by comparing the forward modeled summary statistic to the same statistic from an observed dataset (e.g., the data histogram or mean, etc.). This model-to-data comparison can be done in different ways and a typical approach is rejection, where any parameter set that produces a summary statistic that differs from the observed data by more than some prespecified threshold is rejected. Recent examples in astronomy where ABC forward modeling has been applied include Type Ia supernova cosmology, weak-lensing peak counts, and galaxy demographics \citep{Cameron2012, Weyant2013,Lin2015}. 

Unlike most ABC use cases where the posteriors of all (or most) of the model parameters are constrained, we choose to focus on $Z_\text{v}$ and treat all of the other known parameters with strong priors. In other words, while our ABC forward-modeling approach enables one to simultaneously constrain all of the parameters that go into the observed $v_\text{esc}$ profile, we choose to focus only on $Z_\text{v}$. 

For instance, we could define a grid of values for all of the required parameters that produce a projected phase space including the potential shape parameters, the cosmological parameters, the number of galaxies in the projected phase space, $Z_\text{v}$, as well as the parameters describing the distributions of the galaxy orbits. Given this forward map, we could quantify the $n$-dimensional posterior of those parameters for an observed galaxy cluster by keeping all allowable combinations where the modeled projected phase-space edge matches the observed projected phase-space edge. We could also jointly constrain the phase-space data to the projected density profile as well as the projected velocity dispersion profile. We plan to investigate this generalized approach in a future work. For now, we focus solely on a single parameter: $Z_\text{v}$. Our aim is therefore simplified to address how well $Z_\text{v}$ can be characterized in a constrained parameter space.

\begin{figure}[t]
\centering
\includegraphics[width=1.\linewidth]{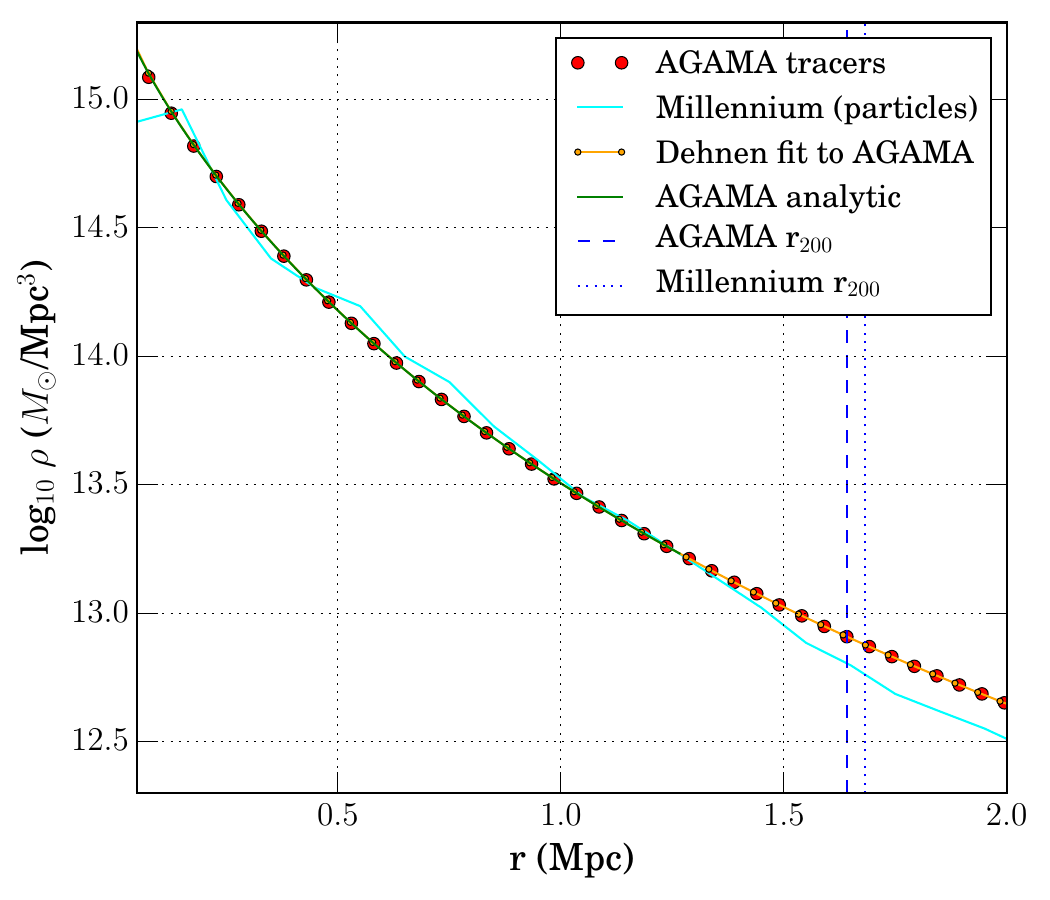}
\includegraphics[width=1.\linewidth]{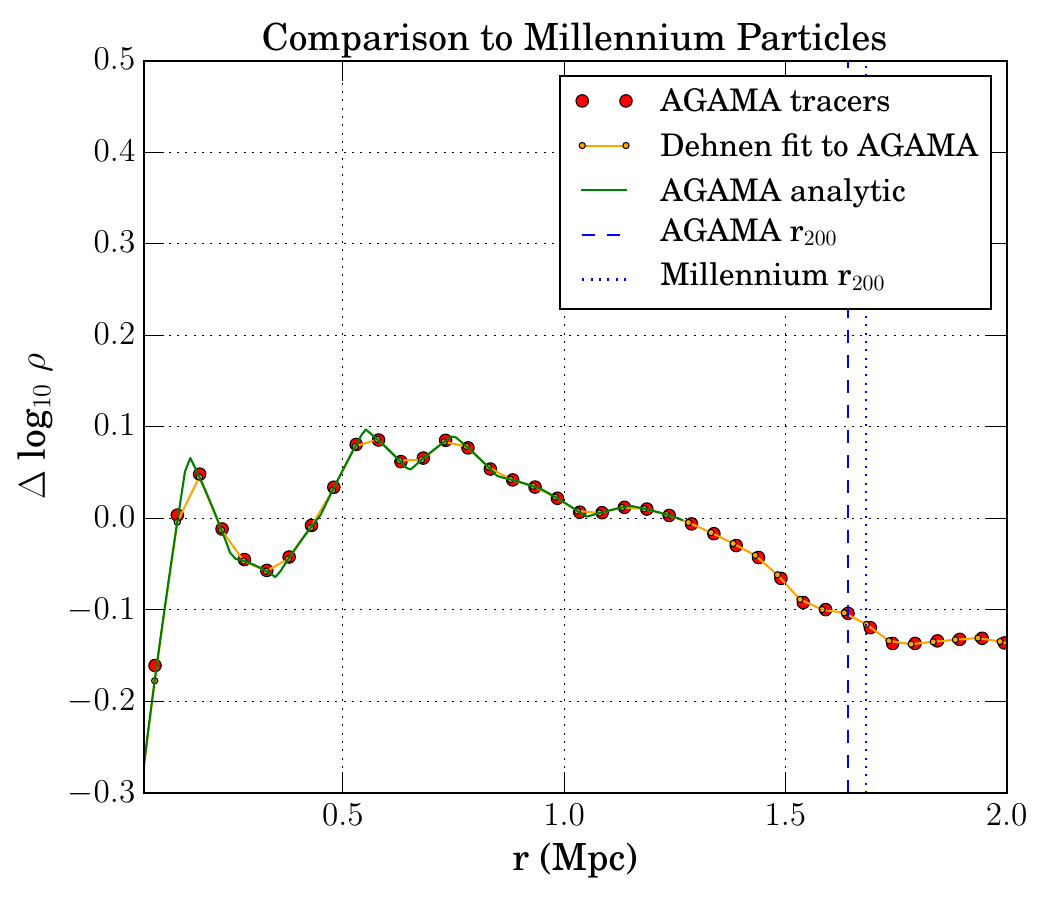}
\caption{Density profile for a single cluster in the Millennium halo sample (ID: 109000144053659, $M_{200} = 4.04 \times 10^{14} M_\Sun$, $r_{200} = 1.46 \, \text{Mpc}$). This density profile is measured from the particles and then fit with a Dehnen profile (Equation (\ref{eq:gamma_den})). These fit parameters ($M= 1.11 \times 10^{14} M_\Sun$, $r_0=1.12 \, \text{Mpc}$, $n = 1.19$) are then used to generate a mock AGAMA phase space based on the density and assuming $\beta = 0$. Then, the AGAMA density profile is measured (orange) from the phase-space tracer data (red) and compared to the AGAMA analytical expectation (green). Also, the location of $R_{200}$ is shown from the simulation and from the AGAMA phase-space data. For clarity, the log$_{10}$ difference is compared to the particles in the lower panel.
}
\label{fig:density_profile}
\end{figure}


\subsection{Phase-space Algorithm}

\subsubsection{Step \#1: Characterize the Sample Inputs}
We begin by defining an example cluster with the following {\it a priori} known constraints:
\begin{enumerate}
    \item The cosmology ($\Omega_{\Lambda}$, $H_0$ in a flat universe)..
    \item The parameters that describe the radially symmetric matter density distribution ($\rho_\text{w}$).
    \item The number of galaxies in the projected phase space in the area $0.3 \, r_{200} < r_\perp < r_{200}$. The symbol $N$ is used throughout this work to refer to this quantity.
\end{enumerate}
Given the above information, we then use the AGAMA framework to generate phase spaces for clusters characterized by their density profiles and their $N$.

\subsubsection{Step \#2: Density Profiles}
There exist analytic formulae that have been shown to fit the density profiles of halos in $N$-body simulations. We use the Dehnen profile \citep{Dehnen93} and solve the Poisson equation to have an analytic representation of the potential in a noncosmological context: 
\begin{subequations}\label{dens_dehnen}
         \begin{align}
          \rho(r) &=  \frac{(3-n) {\rm M}}{{\rm 4\pi}}\frac{r_0}{r^n}\frac{1}{(r+r_0)^{4-n}} \label{eq:gamma_den} \\
          \phi(r) &= -\frac{{\rm GM}}{r_0}\frac{1}{2-n}\Big{[}1-\Big{(}\frac{r}{r+r_0}\Big{)}^{2-n}\Big{]},  n \ne 2 \label{eq:gamma_pot} \\
          &= \frac{{\rm GM}}{r_0}{\rm ln}\frac{r}{r+r_0},  n=2. \nonumber
         \end{align}
\end{subequations}
We can then use Equation (\ref{v_esc_full}) to build an analytic representation of the escape-velocity profile given the density fit parameters $r_0$, $M$, and $n$ as well as the cosmological parameters via Equation (\ref{eq:gamma_den}). An example Dehnen fit to a density profile measured on the particles in the Millennium simulation is shown in Figure \ref{fig:density_profile} (top). We note that it is now established that massive halos have significantly steeper outer density profiles than a classic Navarro, Frenk, and White model \citep{Navarro:1995iw, Diemer:2014gba, Miller:2016fku}.
        
\subsubsection{Step \#3: AGAMA implementation}\label{subsection:agama_implement}
AGAMA  (Action-based Galaxy Modelling Architecture) is a software library that offers a wide range of functionality for dynamical studies of gravitational systems in a noncosmological context \citep{AGAMA}. For this work, we use AGAMA to generate six-dimensional phase spaces for spherically symmetric galaxy clusters. We use the Cuddeford–Osipkov–Merritt model \citep{1979PAZh....5...77O, 1985AJ.....90.1027M, 1991MNRAS.253..414C} for a spherically anisotropic form of the distribution function with anisotropy based on the functional form $\beta(r) = [\beta_0 + (r/r_a)^2]/[1 + (r/r_a)^2]$ (if $r_\text{a} < \infty$, the anisotropy coefficient tends to 1 at large $r$ (Osipkov–Merritt profile), otherwise it stays equal to $\beta_0$ everywhere and the models with constant $\beta$ are found by setting $r_\text{a} = \infty$). This is described in Appendix 6.1 and Section 2.5.3 of \citet{Vasiliev2018}. We then draw positions and velocities from a physically realistic dynamical system for a given Dehnen-based density profile and constant (prespecified) velocity anisotropy $\beta$. Unless otherwise stated, we use $\beta = 0$ (isotropy) as our fiducial value, and we test whether this choice affects the measured projected suppression of the escape edge.

\begin{figure}[t]
\centering
\includegraphics[width=0.99\linewidth]{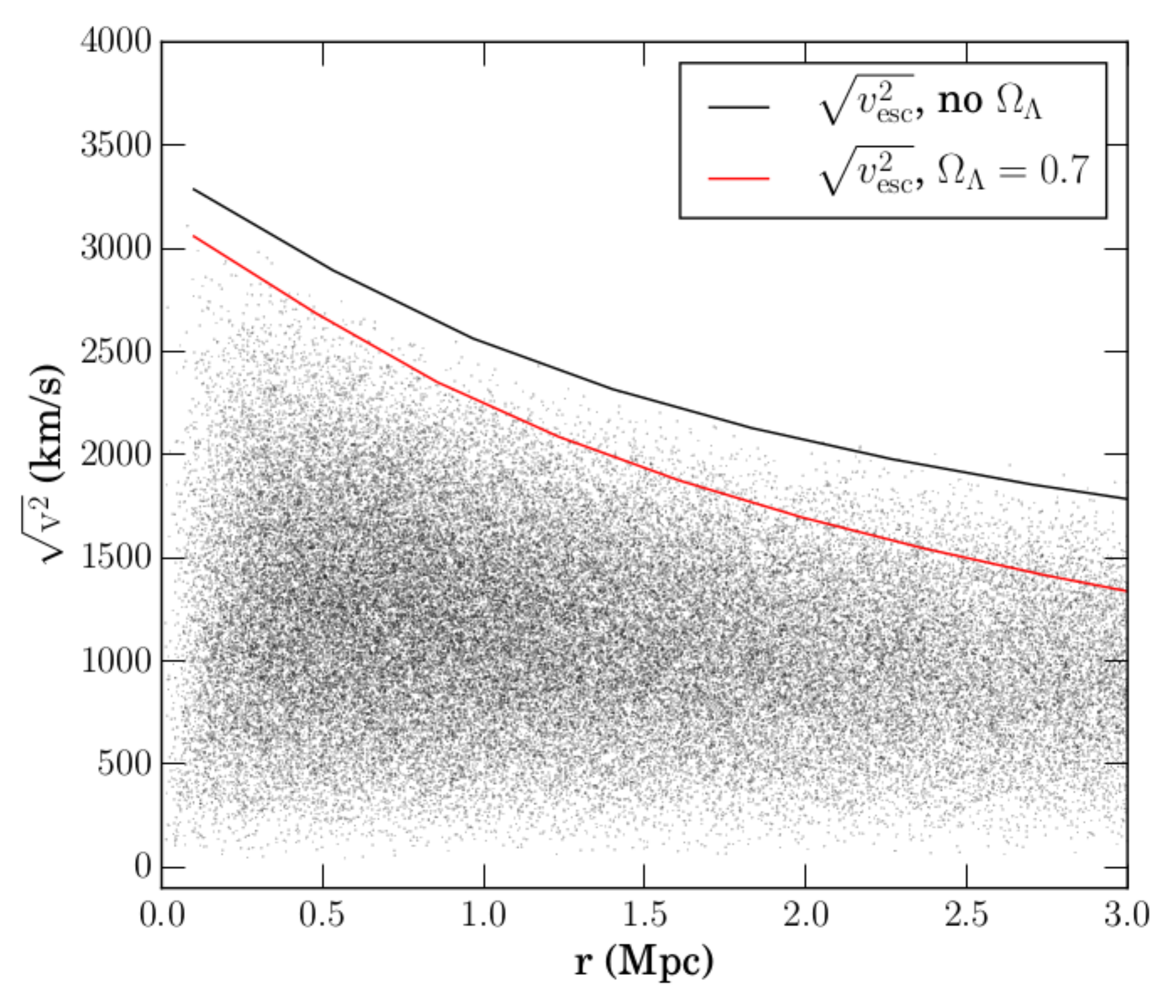}
\caption{3D phase space generated with AGAMA (the same parameters as in Figure \ref{fig:density_profile} are used). The square root of the squared 3D velocity is plotted. The escape profile without (black) and with (red) $\Omega_{\Lambda}$ is shown. For the AGAMA cluster phase-space realizations, tracers above the cosmological escape speed (above the red line) are removed before measuring the projected edges and calculating the suppression $Z_\text{v}$.
}
\label{fig:escape_cull}
\end{figure}

\subsubsection{Step \#4: Culling Escaped tracers}
As noted above, AGAMA is designed to work in a noncosmological context. However, the real universe (and the simulations we will test against) is in a $\Lambda$CDM cosmological background. The effect of the accelerating spacetime on the phase-space data is discussed in Section \ref{section:motiv}.  In a universe with a cosmological constant, galaxies traveling along radial orbits, perpendicular to the line of sight, and above the escape speed would reach the virial radius of the cluster in $\sim$500 Myr. For galaxies above the escape speed but on radial orbits aligned with the line of sight, the current expansion rate (i.e. Hubble flow) would increase their velocity relative to the cluster to $> 100$km/s above the escape edge on a similar timescale (where we assume a virial radius of $1.5 \, \text{Mpc}$).  In $N$-body simulations, these tracers naturally escape and can be cleanly separated using the phase-space data \citep{Behroozi2013,Miller:2016fku}. To incorporate cosmology onto the AGAMA phase-space data, we remove all tracers that have a 3D velocity that is higher than the cosmological escape speed given by Equation (\ref{v_esc_full}). We illustrate this step in Figure \ref{fig:escape_cull}.

\subsubsection{Step \#5: Line-of-sight Projection}
After we cull these tracers, we project along lines of sight from a distance of $30 \, \text{Mpc}$. We follow the same procedure as described in \citet{Gifford:2013ufa} to build the projections, and we treat the viewing angle as a random variable along the $z$-axis. We then calculate the projected phase-space escape edges as described in \citet{Gifford:2013ufa} and \citet{Gifford2017}. Note that we generate the AGAMA phase-space data out to $10 \, \text{Mpc}$. Therefore, the projected phase spaces have some interlopers. A more realistic treatment of interlopers would come from $N$-body simulations.

\begin{figure}[t]
\centering
\includegraphics[width=1.\linewidth]{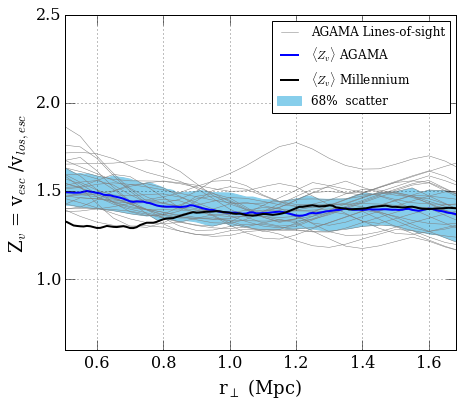}
\caption{Radial profiles of escape-velocity suppression: comparison of predictions from an AGAMA modeled cluster to a halo in the Millennium simulation. This example is for HaloID = 34010484000003 with $m_{200} = 4.52\times10^{14}M_{\odot}$. The two clusters have the same density profile we sample number N. The thin blue lines are the velocity ratio ($Z_\text{v} = v_\text{esc}/v_\text{esc,los}$) of 30 lines of sight to the AGAMA cluster. The  thick blue line and blue shaded region are the mean and $68\%$ scatter. The thick black line is the mean $Z_\text{v}$ of the the Millennium cluster. The $x$-axis covers the radial range $0.3 \, r_{200} \le r \le 1.5\, r_{200}$. 
\centerline{}}
\label{fig:second_approach_step}
\end{figure}

\begin{figure*}
\plottwo{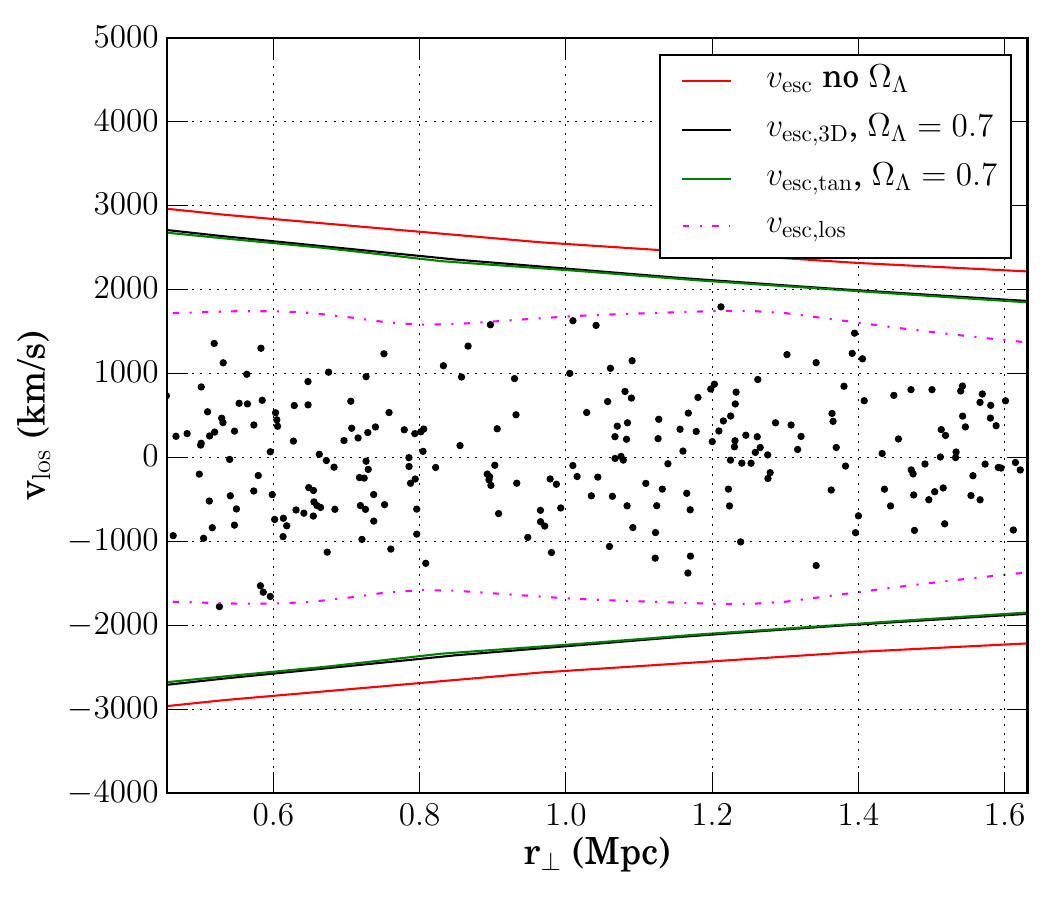}{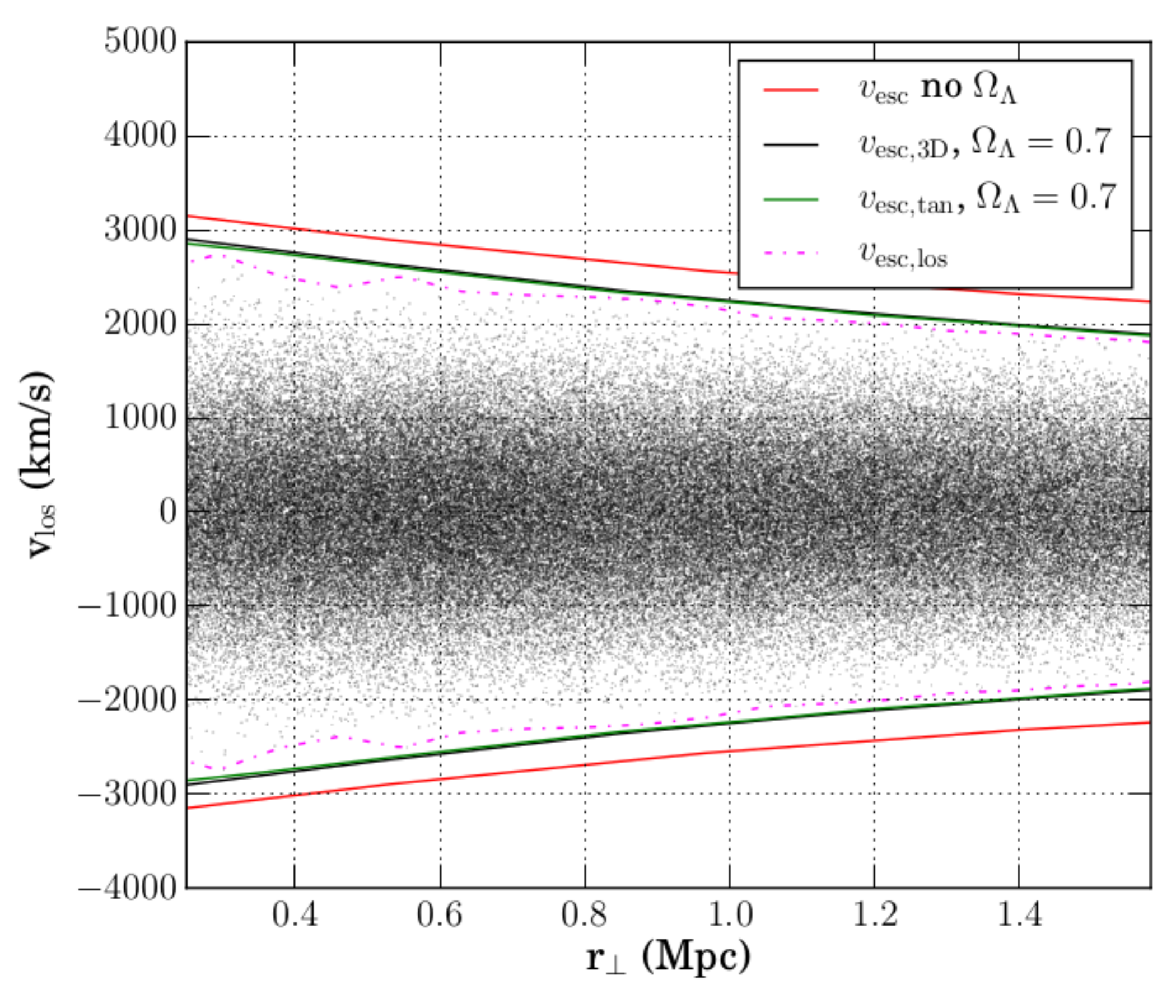}
\caption{A mock AGAMA cluster line-of-sight projected phase space is generated as described in the text. On the left, a few hundred tracers in the phase space are sampled. On the right, $\mathcal{O}(10^5)$ tracers are used. Also, the measured projected edge (dashed line), which is clearly more suppressed at low sampling, is shown. The suppression is measured relative to the cosmological 3D escape edge from Equation (\ref{v_esc_full}) (black). Also, the 3D escape-velocity profile as measured using just the tangential component of the velocity vector for tracers at the edge using $\mathcal{O}(10^6)$ tracers (green) is shown. Given enough sampling, the 3D escape edge is observable in projected data, and suppression is purely statistical. }
\label{fig:example_agama_phase}
\end{figure*}

\subsection{Phase-space Realizations}
Based on steps \#1-5 above, we can create any number of cluster phase-space realizations through this forward modeling. However, we need a set of cluster density profiles to build the model phase spaces.
There are a few options that we could employ to define the parameter values for the cluster phase space we wish to forward model. We could use real data such as the SDSS-C4 sample \citep{Miller2005}. We could use a Jeans analysis of the density and projected dispersion profile \citep{Stark:2017vfa}. However, our choice is to use a cluster sample based on the Millennium $N$-body simulation. This allows us to quantitatively assess realistic effects like nonsphericity, hyper-escape-speed galaxies, and interlopers. We want to stress that we are not calibrating any free parameter in our model to this simulation. The Millennium halos simply provide a representative cluster sample with density profiles, sampling rates, and 3D and projected tracer velocities, all within a fixed and known cosmology.

We use the sample of 100 clusters defined in \citet{Gifford:2013ufa}, which are all below $z= 0.15$, similar to the depth of the SDSS main spectroscopic sample. We extract an average projected profile for each cluster based on 100 random lines of sight within a $60h^{-1} \, \text{Mpc}$ box. These simulated data stem from the Millennium $N$-body simulation \citep{springel2005millenium}. \footnote{The Millennium $N$-body simulation was done with $\Omega_\Lambda=0.75$, which is higher than the value of $\Omega_\Lambda=0.685$ inferred from the \cite{Aghanim:2018eyx} cosmic microwave background data}. Particles from these simulations are used to calculate a Dehnen mass-density profiles (Equation (\ref{eq:gamma_den})) which can be used to also calculate the radial escape profile from Equation (\ref{eq:gamma_pot}) and Equation (\ref{v_esc_full}). 

The cluster masses in this sample are widely spread ($9.3 \times 10^{13} - 1.03 \times 10^{15} M_\Sun$) with the average mass $\langle M \rangle =2.34 \times 10^{14} M_\Sun$ and $\langle R_{200} \rangle =0.95 \,  \text{Mpc}$. We show an example density profile fit in Figure \ref{fig:density_profile}. Note that a full statistical characterization of the Dehnen profile fits to these systems is presented in \citet{Miller:2016fku}. The accuracy and precision are generally quite good over the virial region, as shown in the example cluster in Figure \ref{fig:density_profile}.

Given the density fits, a known cosmology, and a specified tracer sampling rate, we can create projected phase-space realizations using steps 1-5. We then characterize the suppression function $Z_\text{v}$ as the ratio of the underlying radial escape profile to the subsampled and projected phase-space profile edge. 

Note that we can also do this directly on the $N$-body simulation data. 
We use both the particles and the semianalytic galaxies from \citet{guo}. The use of the semianalytic galaxies limits the maximum limit of the phase-space sampling.
To cover a typical range of the number of phase-space galaxies per cluster ($N$) as expected for real data, we create subsets of projected galaxy positions and velocities for the projected galaxies in the simulated halos by varying the apparent magnitude limits. The semianalytic galaxy dataset with the bright magnitude limit provides clusters with the number of galaxies in the projected phase space from  $19<N_\text{l}<257$ with the average number $ \langle N_\text{l} \rangle =58$, while the deeper dataset contains around twice as many galaxies per cluster as the set $N_\text{l}$: $40 < N_\text{h} < 525$ with the average $\langle N_\text{h} \rangle =118$. Note these sets are different descriptions of the same halos, with the only difference being a higher number of dimmer and less massive galaxies per cluster. 

In Figure \ref{fig:second_approach_step} we present an analysis that compares our modeled suppression for $30$ lines of sight to a single cluster. The median and $68\%$ scatter around the median are shown as the blue band. In this figure, we defined the cluster parameters from a specific halo in the Millennium simulation for which we also measure $Z_\text{v}$ using a set of semianalytic galaxy positions and velocities (see Section \ref{results}). Because the Millennium simulation contains the cosmological acceleration, we do not alter the simulation phase-space data. The projections and the escape surfaces are otherwise calculated identically to the AGAMA tracers, for which we match to the number of semianalytic galaxies in the simulation halo.  We find that the suppression quantified from the forward model matches the suppression from the $N$-body simulation (black). We conclude that our model is working and that the realistic treatment of interlopers in the simulation data is not a significant contributing factor to the model.

\begin{figure*}
\includegraphics[width=0.33\linewidth]{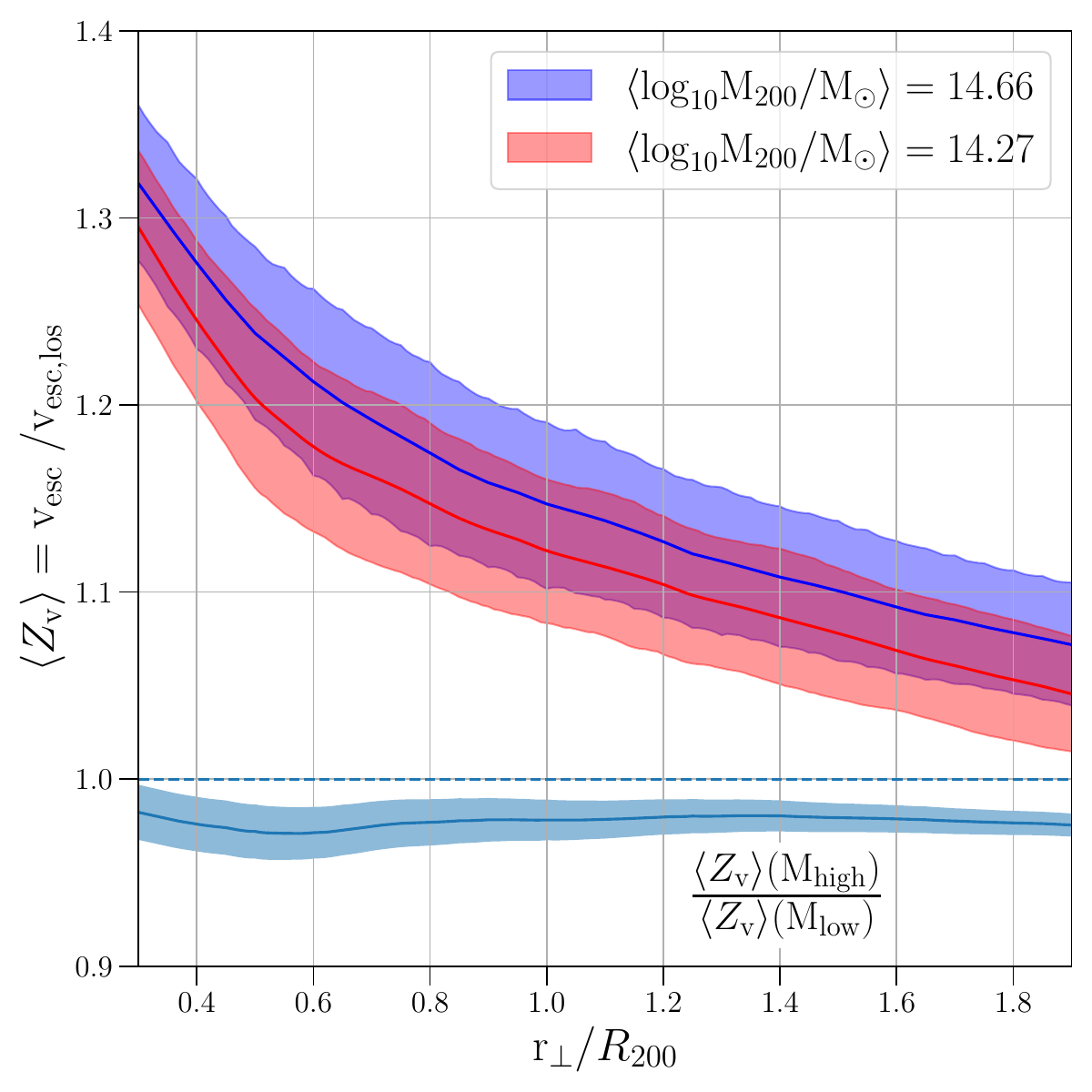}
\includegraphics[width=0.33\linewidth]{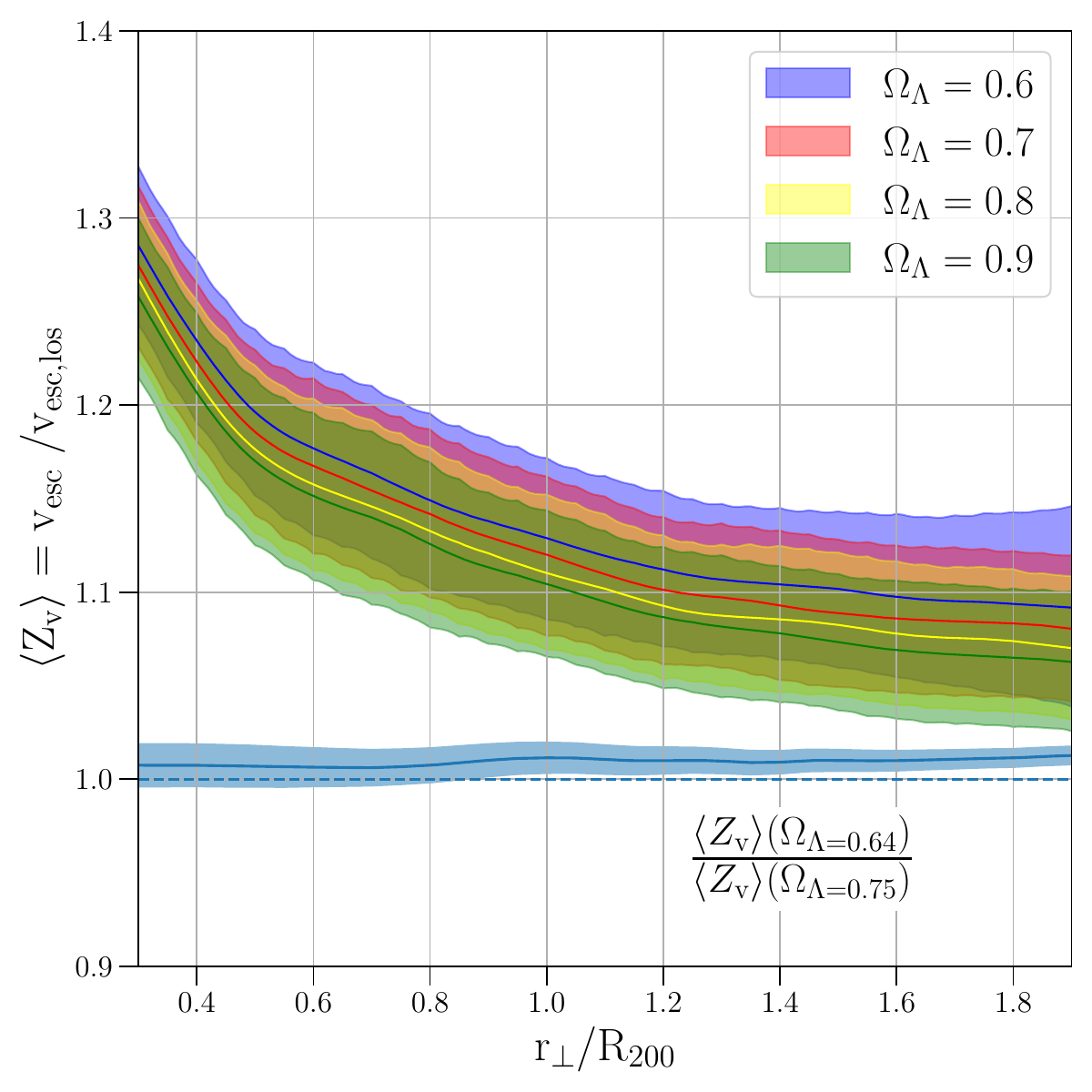}
\includegraphics[width=0.33\linewidth]{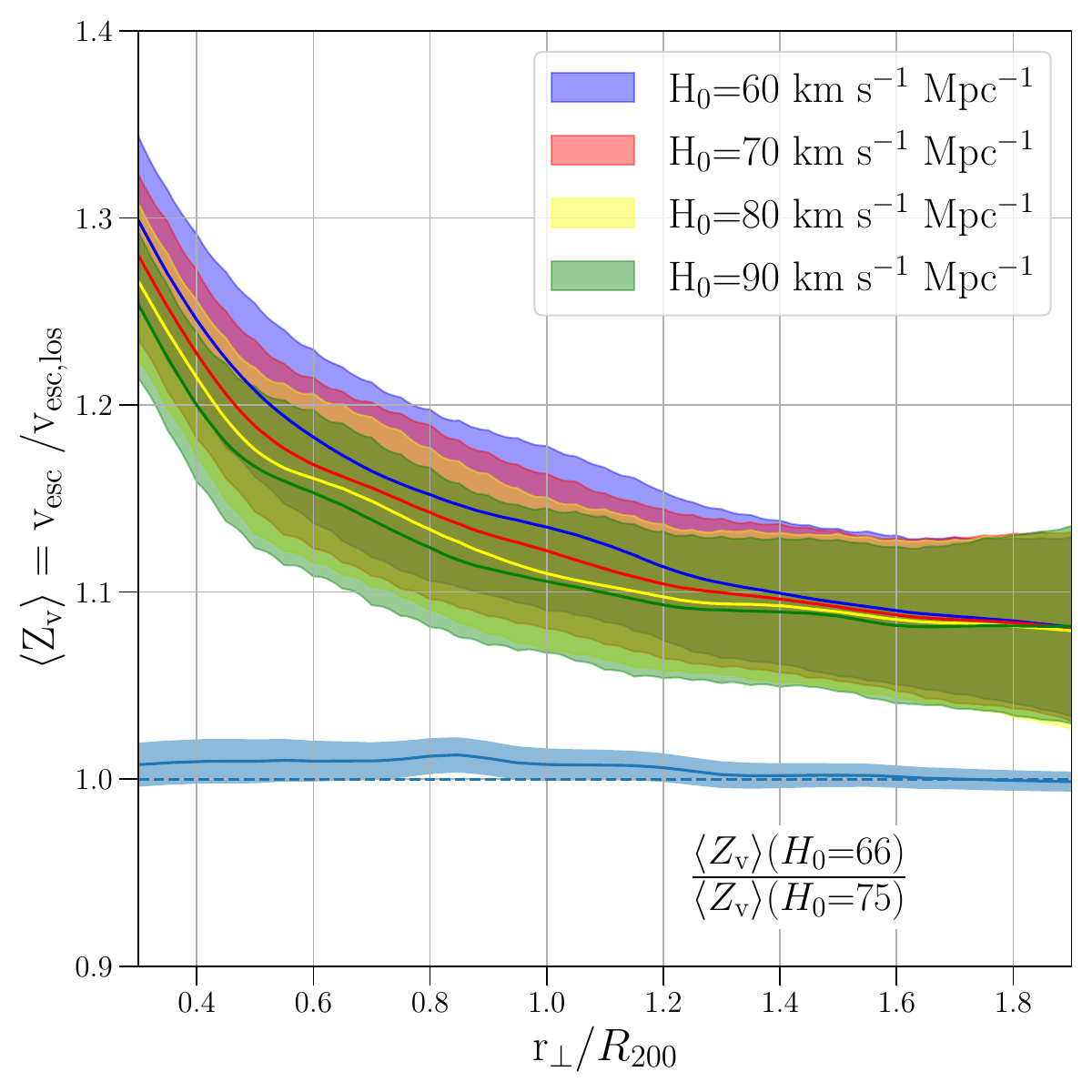}
\caption{Radial profiles of the escape-velocity suppression $\langle Z_\text{v} \rangle$: effects of cluster mass, dark energy parameter, and Hubble constant. AGAMA-generated phase spaces sampled with $N=4837$ tracers. The measurement over 100 lines of sight is averaged, shown as the solid lines. The colored bands represent the range containing 68\% of the scatter in the measurement.  The left panel splits the 100 cluster sample into two mass bins. The middle and right panels use the full set of 100 clusters. The bottom light blue band shows the ratio of the $Z_{\rm v}$ for the parameter as indicated. The width of this band is the 1$\sigma$ standard deviation on the ratio using the errors on the mean $Z_{\rm v}$.
}
\label{fig:change_all_500}
\end{figure*}


The analytical approach enabled by the AGAMA framework allows us to systematically test the suppression function against simulations and in controlled environments, where we can create multiple realizations. For instance, Figure \ref{fig:example_agama_phase} shows example projected phase spaces for the same cluster with different samplings. From this figure we can see how the suppression is apparent in the low-sampled system, but almost nonexistent in the (unrealistic) highly sampled system.

%% file: results.tex
From here on we describe the algorithm defined in the previous section as our ``analytical model.'' This is because it is based purely on an analytic description of the distribution function of precessed orbits in an extended mass profile and in a cosmological background.

\subsection{The Dependence of $Z_\text{v}$ on Cosmology, Mass, and Velocity Anisotropy}\label{subsection:cosm_indep}
Given some starting parameters that allow us to measure $Z_\text{v}$, we ask whether that measurement is sensitive to changes in those initial parameters. We now test whether the suppression depends on the underlying mass of the cluster, the cosmology, or the velocity anisotropy. 

Recall that in order to measure $Z_\text{v}$, we are required to define a cluster through its density profile and the number of galaxies in the projected phase space $N$. Even if we do not require a precise match between the predefined mass/density profile to the modeled system, we still need some starting point to build the phase space. So, we rephrase this new test in such a way as to ask whether the ratio of 3D escape to a projected profile has any quantifiable dependence on the underlying cluster total mass, the cosmology, or the velocity anisotropy. 

Imagine the scenario where a weak-lensing mass profile is made available and followed up with spectroscopy to produce $\sim$100 or so galaxies in the range $0.3 \le r_\perp/R_{200} \le 1$. In practice and given the correct underlying cosmology, the weak-lensing-based prediction of the escape edge and the measured escape edge should agree (to within some degree of scatter), with the only free parameter being the suppression due to the undersampling of the projected phase-space data. However, we want to be sure that the suppression term we infer from our analytical model is unbiased, regardless of the input weak-lensing mass to the model. This is because the weak-lensing mass could in fact be wrong. If the suppression term is independent of the underlying cluster mass and cosmology, then the escape-profile-based mass becomes a powerful tool to characterize weak-lensing systematics (or cosmology, which could also be varied).

In order to quantify the smallest possible dependencies, we use our highest sampled phase spaces with $N=4837$, which is well beyond what could be achieved observationally. For this analysis, we also increase our line-of-sight sampling to 100 unique views. We tested the statistical normality of line-of-sight $Z_{\rm v}$ distributions and confirm that they are Gaussian, justifying the use of means and standard deviations to interpret the significance of any dependencies.

\subsubsection{The Dependence on Mass}
Recall that our predefined cluster density profiles cover a wide range of masses (see Section \ref{approach_desription}). We divide our sample of 100 clusters into a high- and low-mass subsets. We then measure the suppression as a function of radius. In Figure \ref{fig:change_all_500} (left), we show $Z_{\rm v}$ averaged over 100 lines of sight and over the 50 clusters in each subset. We plot the mean values as well as the 16th and 84th percentiles from the 50 clusters in each high- and low-mass subset. 

The bottom band near unity in Figure \ref{fig:change_all_500} (left) is the ratio of the means of the high-mass and low-mass suppression profiles and its combined error on those means. We then take the radial average over the range of interest ($0.3 \le r/R_{200} \le 1$) and find $0.981 \pm{0.003}$ with no statistically significant dependence on radius. We hypothesize that this small variation in $Z_{\rm v}$ as a function of cluster mass may be a result of holding the number of phase-space tracers fixed as opposed to holding the density of tracers fixed (i.e., working in terms of $R_{200}$ reduces the number of tracers per radial bin for the high-mass subset in comparison to the low-mass subset). Because the dependence is so small compared to the suppression itself, we do not investigate further.

\subsubsection{The Dependence on Cosmology}\label{subsection:dep_cosm}
We can also test whether cosmology plays a role in the characterization of the suppression function. This would be difficult using the $N$-body simulations, which rarely cover a wide range of cosmological parameters. Because the AGAMA framework is noncosmological, we can choose a variety of values of the underlying cosmological parameters to cull the escaped galaxies (see Figure \ref{fig:escape_cull}). We vary the Hubble constant ($H_0 = 60, 70, 80, \text{and} \, 90$ [$\text{km s}^{-1} \text{Mpc}^{-1}$]) and the energy density of the dark energy ($\Omega_{\Lambda} = 0.6, 0.7, 0.8, \text{and} \, 0.9$) in our flat $\Lambda$CDM cosmology and remeasure the suppression function in Figure \ref{fig:change_all_500} (middle, right).

As with mass, we find a small dependence on $\Omega_{\Lambda}$ as shown in Figure \ref{fig:change_all_500} (middle). We plot the ratio of two of the mean $Z_{\rm v}$s and its error as the band near unity. To plot this ratio, the widest upper and lower bounds of the observed $\Omega_{\Lambda}$ were used \citep{Aghanim:2018eyx, DESY1}. For this limit, the ratio averaged over the range of interest ($0.3 \le r/R_{200} \le 1$) is 1.008 $\pm{0.002}$ with no statistically significant radial dependence.

We conduct the same analysis for when we vary the Hubble constant and show the results in Figure \ref{fig:change_all_500} (right). The bottom band is the ratio of two of the mean $Z_{\rm v}$s. We show this ratio for the widest observed $H_0$ range based on current high- and low-redshift measurements (see, e.g., \cite{H0}). The ratio averaged over the range of interest ($0.3 \le r/R_{200} \le 1$) is 0.997 $\pm{0.002}$ with no statistically significant radial dependence.

\begin{figure}[t]
\includegraphics[width=0.93\linewidth]{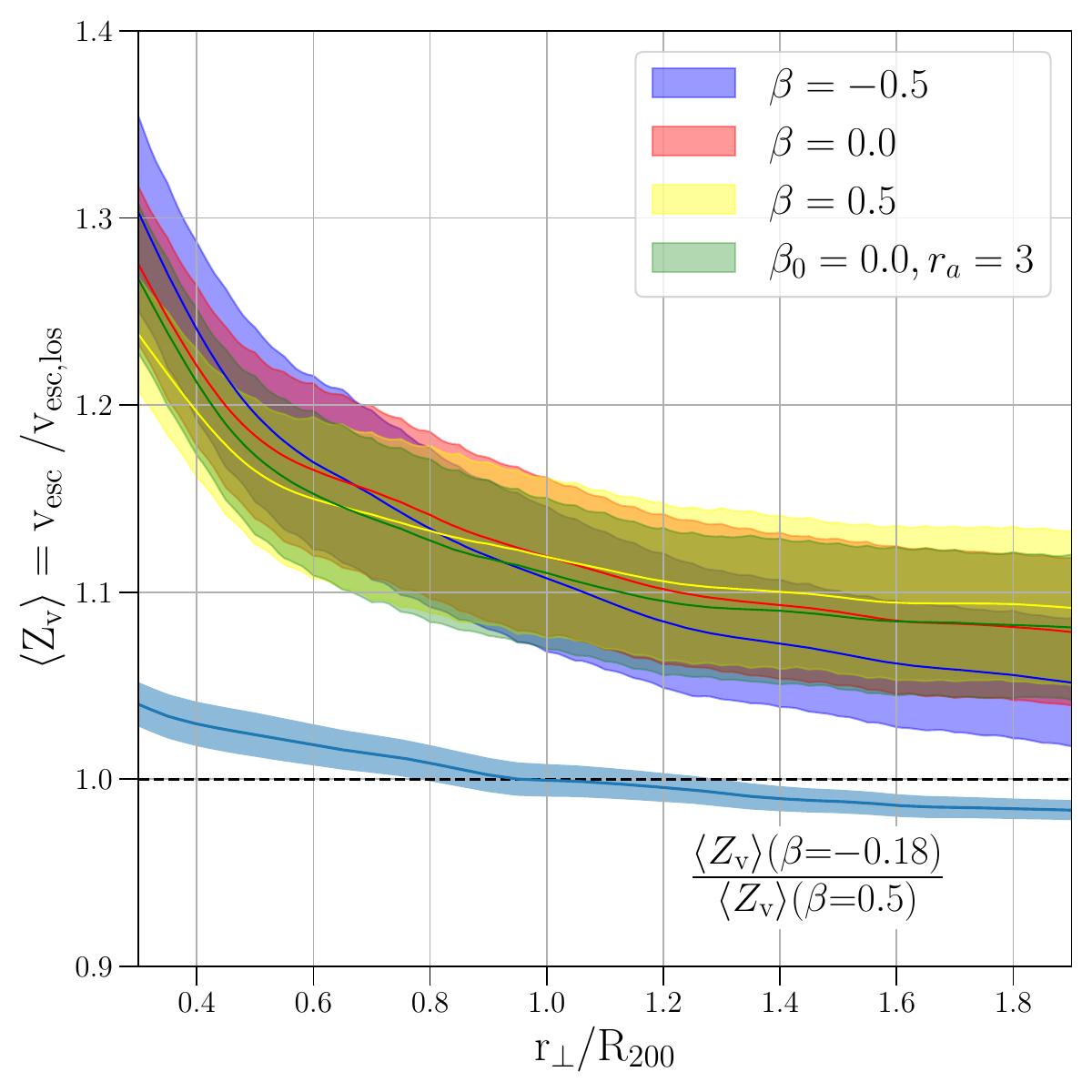}
\caption{The same as Figure \ref{fig:change_all_500} except we vary the anisotropy parameter $\beta$. No dependence of $\langle Z_\text{v} \rangle$ on $\beta$ is found (see Section \ref{subsection:dep_beta}).
}
\label{fig:different_beta}
\end{figure}

\subsubsection{The Dependence of $Z_\text{v}$ on Velocity Anisotropy}\label{subsection:dep_beta}

\cite{Diaferio:1999wg} introduced the approach of connecting $v_\text{esc}$ and $v_\text{esc,los}$ using the anisotropy parameter $\beta(r)$. We test this with our analytical model using the AGAMA framework, where we can control the anisotropy. 

As noted in Section \ref{approach_desription}, our AGAMA modeling so far is done using $\beta = 0$ (isotropic orbits).  The AGAMA Cuddeford–Osipkov–Merritt distribution function model allows for a range of $-0.5 \le \beta \le 1$. This range is wider than what is found in $N$-body simulations and in real data (e.g., see \citet{Stark:2017vfa}). We study $\beta = -0.5, 0, 0.5$ as well as the case $\beta_0=0$ with $r_a=3$ (based on the functional form presented in Section \ref{subsection:agama_implement}) that more closely resembles the Millennium anisotropy profile and remake the AGAMA phase-space data, leaving all of the parameters (e.g., the density fits) fixed. We then measure the suppression ratio $Z_\text{v}$ and show the results in Figure \ref{fig:different_beta}. 

We conduct the same analysis we did for Figure \ref{fig:change_all_500}. The bottom band in Figure \ref{fig:different_beta} is the ratio of two of the mean $Z_{\rm v}$s for $\beta$ from \cite{2010MNRAS.408.2442W} and \cite{Mamon2019}. Unlike the previous parameters, there is a clear radial trend on the dependence of $Z_{\rm  v}$ with $\beta$ when comparing the upper and lower parameter bounds. Within $R_{200}$, the ratio drops from $\sim$1.02 to $\sim$0.98 and then levels off. When averaged over the range of interest ($0.3 \le r/R_{200} \le 1$), we find that the change in the suppression is 1.005 $\pm{0.01}$.

\subsection{Suppression as a Function of Phase-space Sampling}

The analyses and results through this point reinforce the premise of this paper: {\it the suppression of the radial escape edge in projected data is due to statistical sampling alone.} Having searched for $Z_\text{v}$ dependencies on velocity anisotropy, cluster mass, and cosmology and found little to none, we can now characterize the suppression $Z_\text{v}$ simply as a function of the number of phase-space galaxies. 

\begin{figure*}[t]
\centering
\includegraphics[width=0.49\linewidth]{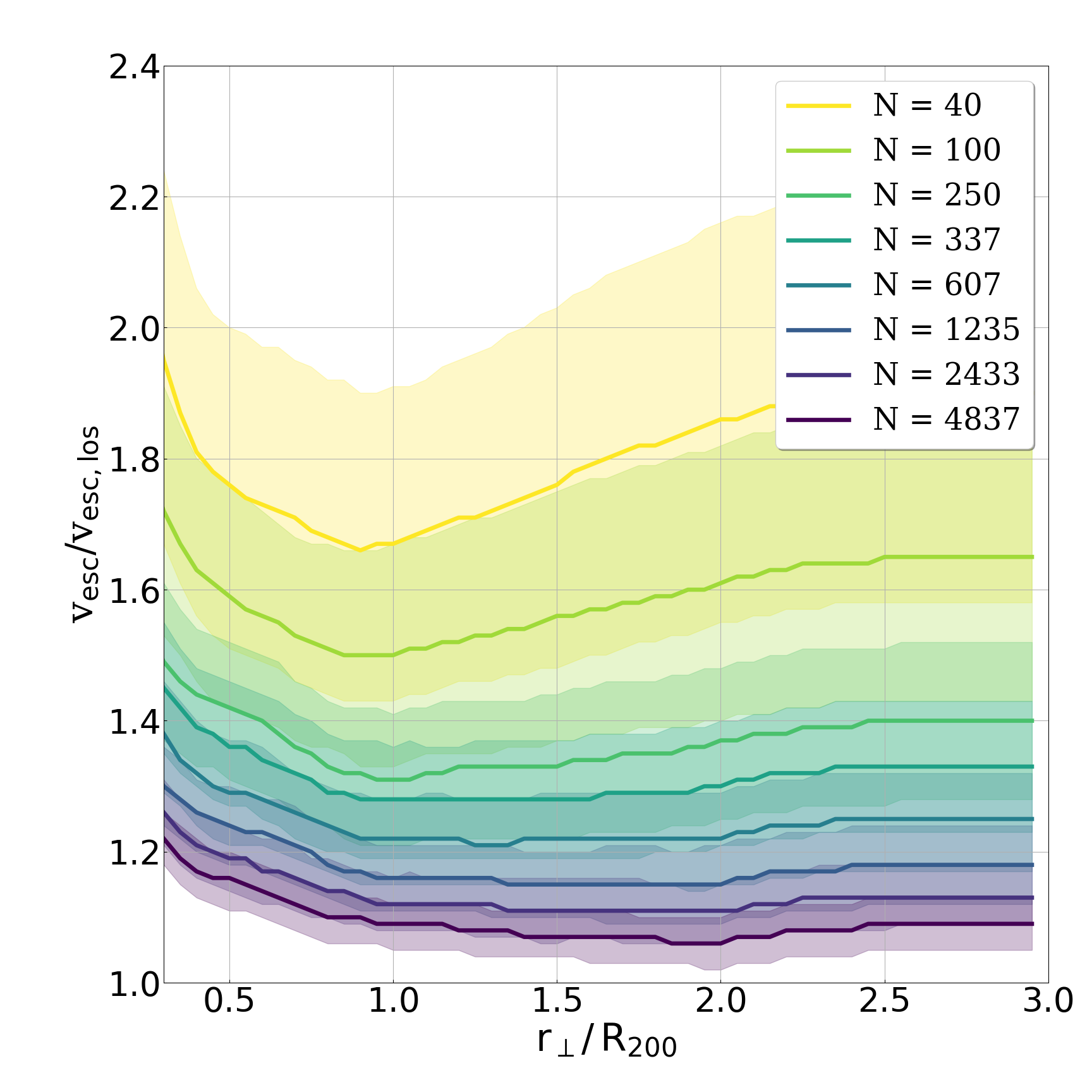}
\includegraphics[width=0.49\linewidth]{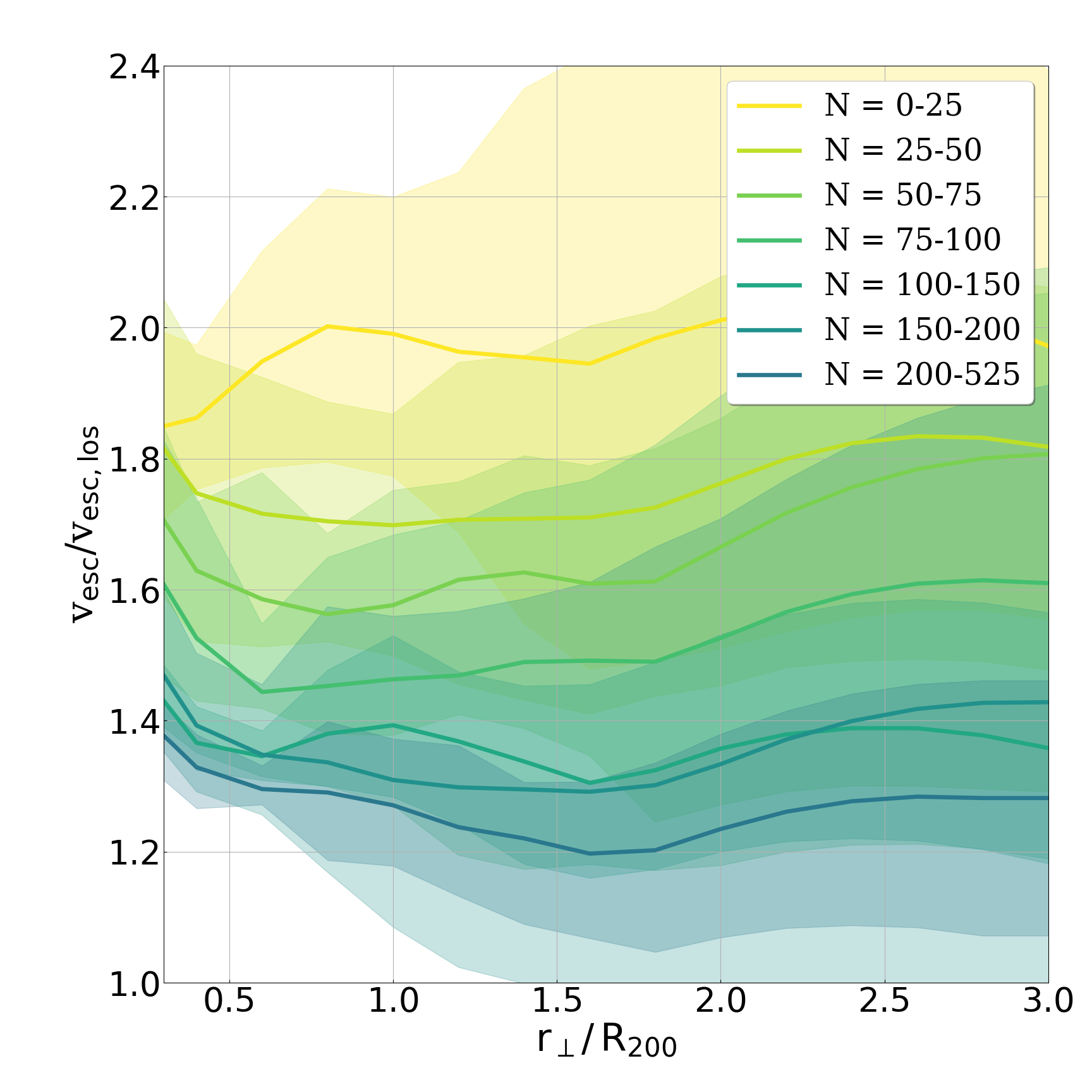}
\caption{The suppression $Z_\text{v}(r)$ (Equation (\ref{ratio_vel})) as a function of the number of galaxies per cluster phase space. {\bf Left:} the predictions from the analytical model. {\bf Right:} the measurement of $Z_\text{v}(r)$ using the semianalytic galaxies from \citet{guo} in the Millennium $N$-body simulation. Thick lines and shaded regions with the same colors are the medians and $68\%$ scatters.}
\label{fig:vel_ratio_millenium}
\end{figure*}

In section \ref{approach_desription} we showed that when using a cluster with a predefined density profile, the phase-space sampling affected how closely we are able to measure the 3D escape edge (see Figure \ref{fig:example_agama_phase}). Our premise is that the suppression value ($Z_\text{v}$) should depend on the number of galaxies in the projected phase space $N$: we predict an increase in $v_\text{esc,los}$ (or a decrease in the projected suppression) as the number of galaxies per cluster increases. In Figure \ref{fig:vel_ratio_millenium} (left), we show this prediction based on the analytical model and by averaging the 100 clusters over 30 lines of sight per cluster and each with a different phase-space sampling $N$. We see that there is a clear dependence between $Z_\text{v}$ ($v_\text{esc}/v_\text{esc,los}$) and $N$. 

We can make the same test using our Millennium clusters. The sample is big enough to split it into six groups based on the number of projected phase-space galaxies $N$: 0-25, 25-50, 50-75, 75-100, 100-150, 150-200, and 200-525. The first four groups are taken from the bright magnitude dataset ($N_\text{l}$), while the last two groups are from the sample with the deeper magnitude limit ($N_\text{h}$). We treat these datasets as being realistic observational data, such that the phase spaces are in principle observable to these magnitude limits with typical astronomical instrumentation. Recall that we are sampling the projected positions and velocities from the \cite{guo} semianalytic galaxy catalogs projected to a distance of $30 \, \text{Mpc}$. Figure  \ref{fig:vel_ratio_millenium} (right) shows that we see the same behavior in the fully evolved simulations as we do in the analytical model. The suppression decreases with increased phase-space sampling. 

\subsection{Quantifying $Z_\text{v}(N)$}\label{subsect:quantif_zv}
\begin{figure}
\centering
\includegraphics[width=0.99\linewidth]{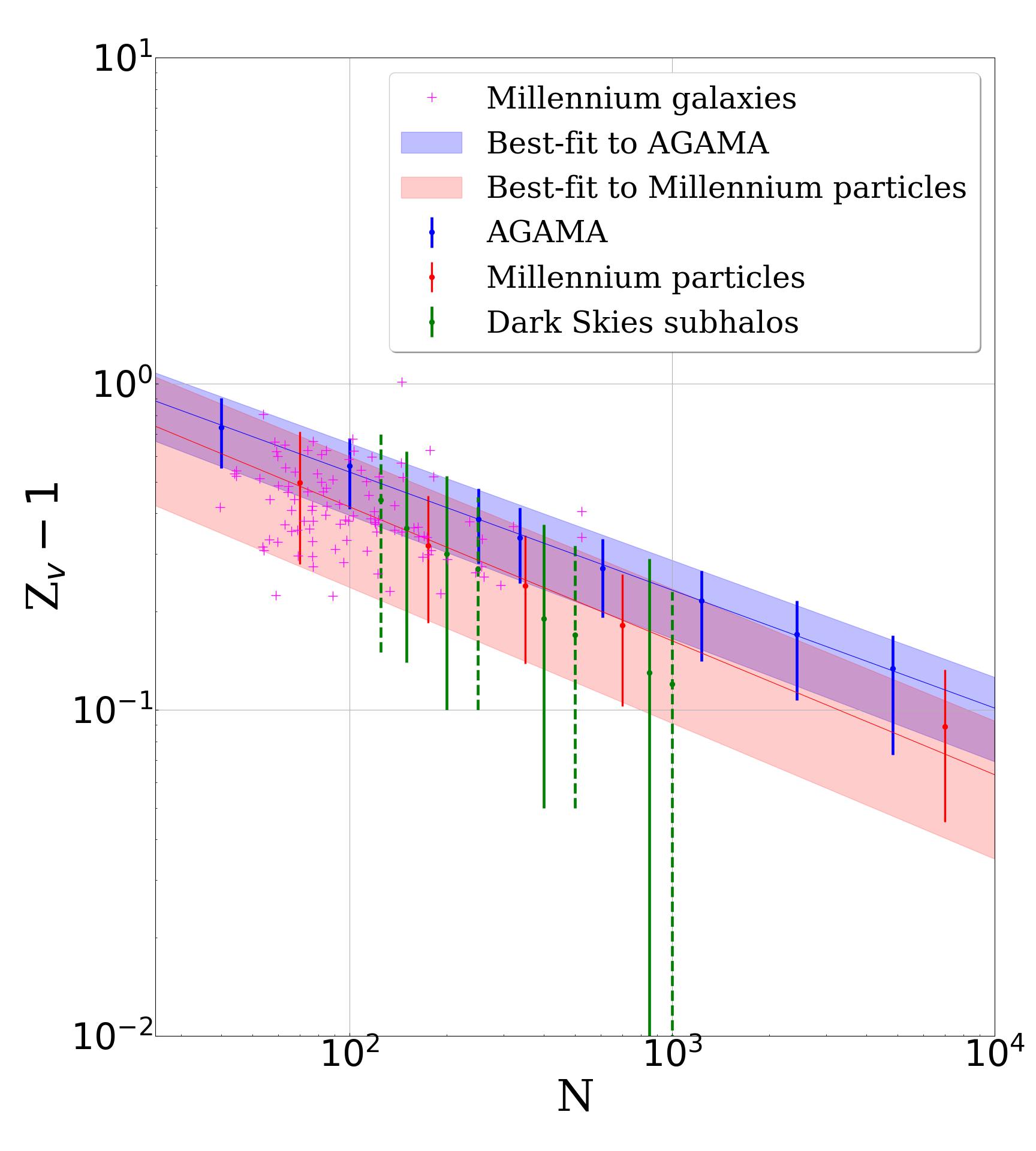}
\caption{The escape-edge suppression $\langle Z_{\text{v}} \rangle - 1$ as a function of the number of tracers, $N$. The mean $Z_{\text{v}}(N)$ are the blue dots, and the bars capture 68\% of the scatter in the data. The one plus power-law fit to the AGAMA results is the thin blue line and the blue band is the area between fits to the AGAMA 16\% and 84\% scatters. Also, the suppression function as calculated on the Millennium halos using both the particles (red dots/bars and 68\% scatter) and the semianalytic galaxies (pink plus signs) is shown. Finally, the measured suppression based on subhalos in the Dark Skies simulations using both massive (dashed green) and less massive (solid green) halos is shown. Good agreement between the simulations and our analytic prediction for $\langle Z_{\text{v}}(N) \rangle$ is found.
}
\label{fig:Z_v}
\end{figure}

We apply our analytical model to create numerous samples of 3D and maximum observed velocity profiles, and we then vary the number of tracers in the modeled projected phase space between $0.3 \le r_\perp/R_{200} \le 1$.  We then calculate the mean $\langle Z_\text{v} \rangle$ over the range $0.3 \le r_\perp/R_{200} \le 1$ and plot it as a function of $N$ in Figure \ref{fig:Z_v}. We also show the 68\% scatter in the data as the blue band. 

We note that in Figure \ref{fig:change_all_500}, the suppression function $Z_\text{v}(r)$ profile shows a slight radial dependence, with a steepening toward the cluster core and in the outskirts, while being flat in between. For the analytic mock clusters, the value of the (negative) slope in the virial region is independent of $N$ for $N > 250$ galaxies (i.e., well-sampled cluster phase spaces). This radial dependence means that the range over which we measure the average value of $Z_\text{v}(r)$ plays a role in its value, and the mean can change by $\sim \pm{5\%}$ for  $N > 250$ when, for example, the radial limit used to measure the mean is varied from $0.5R_{200}$ to $R_{200}$. We also notice similar radial dependencies in the simulations in Figure \ref{fig:vel_ratio_millenium} (right). Possible explanations could include three-body interactions (or a lack thereof) and the cosmological background of galaxies. We leave these to explore in a future effort.

We find that the suppression factor tends toward $1$ at high $N$. With samples as large as $N=10^4$, we would expect to measure a projected escape edge that is only $\sim 10\%$ suppressed compared to the underlying radial escape velocity. However, at low sampling, the edge can be suppressed by as much as a factor of 2. We fit an inverse power law to the suppression $\langle Z_{\text{v}} \rangle$ over the range  $0.3 \le r_\perp/R_{200} \le 1$:
\begin{equation}\label{Z_v_function}
    Z_\text{v}(N) = 1 + \Bigr(\frac{N_0}{N}\Bigr)^\lambda, 
\end{equation}
where $N_0$ and $\lambda$ are the parameters of the model. 
We constrain the fit parameters as $N_0 = 17.818, \lambda= 0.362$. We also measure the cluster-to-cluster scatter as the range on the parameters which contains 68\% of the models. The bottom dashed (16\%) line has $N_0 = 8.533, \lambda= 0.378$, and the upper dashed line (84\%) has $N_0 = 30.989, \lambda= 0.356$. While the ratio $Z_\text{v}$ is presented for the wide range (i.e. $10 \leq N \leq 10^4$), the fitting procedure was done by utilizing only the $40 \leq N \leq 600$ range as this is the typical range of $N$ of the real observed system used in cosmological analyses \citep{Halenka:2018qnj}. Note the fits that we provide here are to the percentiles we plot in Figure \ref{fig:Z_v}. Therefore, these fits are not from a linear regression where the data on the ordinate have error bars. We have not calculated error bars for our estimates of the 16th, 50th, and 84th percentiles. In this sense, the fits are meant to be exact representations of these percentiles we plot and Figure \ref{fig:Z_v} provides a range of suppression values that are equally probable for a given $N$.

We conduct a comparison using the Millennium simulation. For this test, we use both the semianalytic galaxies and the particles. By doing so we can check for whether velocity bias between the particles and the galaxies plays any role and also measure the suppression for a higher $N$ than any nominal galaxy cluster might allow. In Figure \ref{fig:Z_v}, we find good agreement between the predicted $Z_{\text{v}}(N)$ to that observed in the simulation. The constraints from the Millennium simulation on the fit parameters of the functional form $Z_\text{v}$ (Equation (\ref{Z_v_function})) are $N_0 = 14.656, \lambda= 0.450$ (the bottom 16\% line: $N_0 = 3.772, \lambda= 0.438$ and the top 84\% line: $N_0 = 32.582, \lambda= 0.452$).

\subsection{Alternate Simulation Test and Halo-mass Dependence}

Recall that we used the Millennium simulation to enable us to define realistic density. While we did not calibrate any free parameter to the Millennium in our $Z_\text{v}(N)$ model, it is worth making a blind test against a different simulation. We choose the Dark Skies simulation \citep{Skillman2014}.

We choose the Dark Skies ds14g simulation because it balanced a large-enough box size while nearly matching the Millennium particle mass (i.e., resolution). We specifically chose the simulation containing $4096^{3}$ particles of mass $6.1 \times 10^{8}h^{-1} \, M_\Sun$ in an 8$h^{-1} \, \text{Gpc}$ box. This simulation has a flat cosmology with $\Omega_{\Lambda} = 0.7048$ and $H_0 = 68.81 \, \text{km} \, \text{s}^{-1} \, \text{Mpc}^{-1}$ at $z = 0$, which is the data we utilize. Dark Skies utilizes the 2HOT base code, a tree-based adaptive $N$-body method, as opposed to the Gadget-based code used in Millennium.

Unlike the Millennium simulation, which carries with it a number of semi-analytic galaxy catalogs \citep{Bower06,Bertone07,Delucia07,guo}, the Dark Skies simulation only provides us with subhalos. However, there are many more subhalos than there are galaxies for any realistic halo. For the Millennium semianalytic galaxy sample, we applied an absolute magnitude limit to define the phase-space tracer selection \citep{guo}. For the Dark Skies, we adjust the threshold on the subhalo masses to define how many galaxies populate the phase space. We keep only the most massive subhalos above that threshold. Like in the magnitude thresholding in the Millennium, the subhalo-mass thresholding mimics targeting in a spectroscopic follow-up campaign.

We also divided the Dark Skies cluster sample into two halo-mass bins, with each having approximately 10 systems. The low-mass bin has $\langle M_{200} \rangle = 10^{14.34}M_{\odot}$, which closely matches the Millennium sample described at the beginning of this section. We also created a high-mass sample with $\langle M_{200} \rangle \sim 10^{15}M_{\odot}$. Unlike the Millennium clusters or the low-mass Dark Skies halos, the Dark Skies massive clusters are representative of currently available observed weak-lensing and phase-space data \citep{Stark:2017vfa}.

In Figure \ref{fig:Z_v} we show the results of the measured $Z_\text{v}$ function for the Dark Skies data. The dashed green lines are for the high-mass Dark Skies clusters while the solid lines are for the lower-mass systems. As with the Millennium, we find good agreement with our predictions from the analytically generated phase spaces.  We can also conclude that our fit to $Z_\text{v}(N)$ using the analytical model is not influenced by the use of the Millennium sample for a set of predefined cluster density profiles.
Best-fit parameters with 1$\sigma$ errors of the suppression function (\ref{Z_v_function}) are $N_0 = 13.565 \pm 1.460, \lambda = 0.437 \pm 0.016$ for Millennium particles and $N_0 = 18.647 \pm 1.717, \lambda = 0.371 \pm 0.014$ for the analytical model (we do not provide best-fit parameters for Dark Skies simulations as there is not enough data to produce accurate statistics). Note that these best-fit parameters differ from those presented in Section \ref{subsect:quantif_zv}, as those parameters describe the upper and lower ranges that contain 68\% of the data.

\subsection{Systematic Shift of $Z_\text{v}$}
As we showed in Section \ref{subsection:cosm_indep}, there is little to no radial dependence of $Z_\text{v}$ on cosmology and velocity anisotropy. Additionally, there is only a small indication of variations of $\langle Z_\text{v} \rangle$ with the changes in cosmological parameters and velocity anisotropy. While this analysis was done for the case with $N=4837$ tracers, it is pointed out in Section \ref{subsect:quantif_zv} that the real observational systems used in the cosmological analysis have a smaller number of galaxies ($40 \leq N \leq 600$). In this range of $N$, we found small variations in $\langle Z_\text{v} \rangle$. More specifically, by measuring the average over the range of interest ($40 \leq N \leq 600$) $\langle Z_\text{v} \rangle$ in the range of parameters presented in Figures \ref{fig:change_all_500} and \ref{fig:different_beta}, we found the following maximum average variations:

\begin{itemize}
    \item the energy density of the dark energy: $\langle\langle Z_\text{v} \rangle\rangle _{\Omega_\Lambda = 0.6} - \langle\langle Z_\text{v} \rangle\rangle _{\Omega_\Lambda = 0.9} = 0.037$;
    \item present value of the Hubble parameter: $\langle Z_\text{v} \rangle _{\text{H}_0 = 70} - \langle Z_\text{v} \rangle _{\text{H}_0 = 60} = 0.026$;
    \item anisotropy parameter: $\langle Z_\text{v} \rangle _{\beta = 0} - \langle Z_\text{v} \rangle _{\beta=-0.5} = 0.024$,
\end{itemize}
where the notation $\langle\langle Z_\text{v} \rangle\rangle$ used above means that $Z_\text{v}$ is first averaged over the radial range $0.3 \le r/R_{200} \le 1$ and then it is averaged over the range of the number of galaxies $40 \leq N \leq 600$. Note, the ranges of parameters used in the calculation of the above maximum averaged variations are much wider than what are currently constrained from observations (see Sections \ref{subsection:dep_cosm}, \ref{subsection:dep_beta}).

We can draw a couple of important conclusions from these results. First of all, the maximum variations do not resemble trends in two of the three cases as the the maximum differences of $\langle\langle Z_\text{v} \rangle\rangle$ are between cases $H_0 = 70 \, \text{km} \, \text{s}^{-1} \, \text{Mpc}^{-1}$ and $H_0 = 60 \, \text{km} \, \text{s}^{-1} \, \text{Mpc}^{-1}$ (while the range of explored parameters is $60-90 \, \text{km} \, \text{s}^{-1} \, \text{Mpc}^{-1}$) and between $\beta = 0$ and $\beta = -0.5$ (while the range of explored parameters is $-0.5$ to $0.5$). So, it is not clear if the maximum average variations are due to fluctuations in the data or there is actual functional dependence on cosmology and/or velocity anisotropy. We leave this question to explore in the future efforts, and we treat the above variations as systematic uncertainties.

Our second conclusion is that the changes of $\langle Z_\text{v} \rangle$ due to cosmological parameters and velocity anisotropy are significantly smaller than the change due to the number of galaxies. The biggest individual change of $\langle Z_\text{v} \rangle$ is between cases with $\Omega_\Lambda = 0.6$ and $\Omega_\Lambda = 0.9$, and it is $2.7\%$, while the change of $\langle Z_\text{v} \rangle$ due to the increase in the number of galaxies from $N=40$ to $N=600$ is $36.5\%$. The $Z_\text{v}$ dependence on the number of galaxies is at least $13.7$ times more significant than the dependence on the cosmological parameters, mass, or the velocity anisotropy. We thus treat the suppression $Z_\text{v}$ as a  function of the number of galaxies $N$ with percent-level accuracy limited by systematics from the cosmological parameters, cluster masses, and velocity anisotropy.

%% file: disc_concl.tex
The premise of this paper is to determine the cause of the suppression of the escape-velocity phase-space edge in observed cluster phase spaces. We use the AGAMA software framework to generate mock cluster projected phase spaces \citep{AGAMA}. We then use our modeled phase spaces to directly calculate the suppression of the radial escape-velocity profile under different scenarios (Section \ref{results}). We find that with enough tracers, the underlying escape profile is observable in projection.

We examine the suppression of the observed phase spaces (i.e. projected) with tracer samples $\mathcal{O}(10^2)$ to show that cluster mass, cosmology, and velocity anisotropy play no statistically measurable role in the amount of the edge suppression. Instead, we find that the observed suppression of the escape-velocity profile is due to undersampled phase spaces, modeled by a one plus power-law relation to the number of phase-space galaxies, $Z_\text{v}(N)$. For instance, our model predicts that projected escape profiles with $N= 100$ should be suppressed to $\sim$ 70\% of the true escape velocity. We confirm this prediction on two simulation datasets using particles, semianalytic galaxies, and subhalos as the underlying tracers. If one were able to observe $\mathcal{O}(10^4)$ tracers in a cluster, the observed edge matches the underlying radial escape edge to within 10\%.

We conclude that our analytical cluster phase-space modeling enables observed cluster phase space edges to be ``desuppressed'' into the underlying radial escape profile to $\sim 2 \, r_{200}$. Our analytical model frees the escape-velocity technique from the need to calibrate against simulations. By using the absolute velocity maximum to define the edge, we also remove the need for the velocity dispersion to calibrate an ``edge'' as in previous works. This is important because the dispersion can be biased according to the tracer-type \citep{ Biviano2002,Evrard2008,Gifford:2013ufa,Bayliss2017} 

Finally, comparing the value of the suppression as a function of the number of galaxies, the trends in $Z_\text{v}$ with cosmology, mass, or the velocity anisotropy are highly subdominant (more than a factor of 10 smaller in magnitude). This is an important and significant shift from prior interpretations when using the escape edge to infer cluster masses or cosmology \citep{Stark:2016dxf, Stark2017}. Our work provides clear evidence that given a cosmology, the desuppressed escape profile provides a direct constraint on the mass profile of a galaxy cluster  (see Equation (\ref{v_esc_full})). Similarly, if a mass profile were already available from a nondynamical technique (e.g., via the shear profile/weak lensing), the combination of the escape profile and mass profile provides a direct constraint on the acceleration of space-time through $qH^2$.